\documentclass[a4paper,11pt]{article}
\pdfoutput=1 

\usepackage{jcappub} 

\usepackage[T1]{fontenc} 
\usepackage{placeins}
\usepackage[dvipsnames,svgnames,x11names,hyperref]{xcolor}
\usepackage{array}
\usepackage{graphicx}
\usepackage{booktabs}
\usepackage{slashed}
\RequirePackage{mathrsfs}
\usepackage{dsfont}
\usepackage{hyperref}
\hypersetup{
        unicode = true,
        pdftitle = Note , 
        colorlinks = true,
        linkcolor = Black!30!DodgerBlue4,
        citecolor = Black!30!DodgerBlue4,
        filecolor = Black!30!DodgerBlue4,
        urlcolor = Black!30!DodgerBlue4,
}

\usepackage{booktabs}
\usepackage{cancel} 

\renewcommand{\toprule}{\specialrule{1.5pt}{0em}{1pt} \midrule}
\renewcommand{\bottomrule}{\midrule \specialrule{1.5pt}{1pt}{0em}}

\graphicspath{{./plots/}}

\usepackage{floatrow}
\newfloatcommand{capbtabbox}{table}[][\FBwidth]

\newcommand{\erf}{\mathop{\mathrm{erf}}}

\renewcommand{\epsilon}{\varepsilon}

\renewcommand{\vec}[1]{\boldsymbol{#1}}
\renewcommand{\bar}{\overline}

\newcolumntype{L}{>{$}l<{$}}
\newcolumntype{C}{>{$}c<{$}}

\allowdisplaybreaks

\title{\boldmath Prospects for Dark Matter signal discovery and model selection via timing information in a low-threshold experiment}

\author[a]{Riccardo Catena}
\author[a, b, c]{and Vanessa Zema}

\affiliation[a]{Chalmers University of Technology, Department of Physics, SE-412 96 G\"oteborg, Sweden}
\affiliation[b]{GSSI-Gran Sasso Science Institute, 67100, L'Aquila, Italy}
\affiliation[c]{Max Planck Institute for Physics, 80805, Munich, Germany}
\emailAdd{catena@chalmers.se}
\emailAdd{vanezema@mpp.mpg.de}

\abstract{
In the recent years, many low-threshold dark matter (DM) direct detection experiments have reported the observation of unexplained excesses of events at low energies. Exemplary for these, the experiment CRESST has detected unidentified events below an energy of about 200 eV -- a result hampering the detector performance in the search for GeV-scale DM.
~In this work, we test the impact of nuclear recoil timing information on the potential for DM signal discovery and model selection on a low-threshold experiment limited by the presence of an unidentified background resembling this population of low-energy events.~Among the different targets explored by the CRESST collaboration, here we focus on Al$_2$O$_3$, as a sapphire detector was shown to reach an energy threshold as low as 19.7~eV~\cite{Angloher:2017sxg}.~We test the ability of a low-threshold experiment to discover a signal above a given background, or to reject the spin-independent interaction in favour of a magnetic dipole coupling in terms of $p$-values.~We perform our $p$-value calculations:~1) taking timing information into account; and~2) assuming that the latter is not available.~By comparing the two approaches, we find that under our assumptions timing information has a marginal impact on the potential for DM signal discovery, while provides more significant results for the selection between the two models considered. For the model parameters explored here, we find that the $p$-value for rejecting spin-independent interactions in favour of a magnetic dipole coupling is about 0.11 when the experimental exposure is 460 g$\times$year and smaller (about 0.06) if timing information is available.~The conclusion on the role of timing information remains qualitatively unchanged for exposures as large as 1~kg$\times$5~year.
~At the same time, our results show that a 90\% C.L. rejection of spin-independent interactions in favour of a magnetic dipole coupling is within reach of an upgrade of the CRESST experiment~\cite{CRESST:2015djg}.} 

\begin{document} 
\maketitle
\flushbottom

\section{Introduction}
While the identity of our Universe's invisible mass component, Dark Matter (DM), remains unknown, the experimental search for its microscopic constituents progressed rapidly in recent years~\cite{Bertone:2016nfn}.~In particular, the search for nuclear recoils induced by the non-relativistic scattering of Milky Way DM particles in low-background deep underground detectors -- the so-called DM direct detection technique~\cite{Drukier:1983gj, Goodman:1984dc} -- played, and will continue playing, a central role in this context~\cite{Undagoitia:2015gya}.
~For example, the null result of the operating experiments led to stringent constraints on the 
DM-nucleon and -electron coupling, e.g.~\cite{Zyla:2020zbs}, and 
the effort placed on assessing the expected performance of next-generation direct detection 
experiments has been remarkable in recent years~\cite{Trotta:2006ew,Trotta:2007pg,Cerdeno:2007hn,Alves:2010pt,Catena:2014epa,DelNobile:2015rmp,Sandick:2016zut,Kavanagh:2017hcl}.~
In this assessment, the performance of an experiment 
is measured by 
the exposure it 
must reach to reject the background-only hypothesis for a specific background model, or to reject a particle physics model in favour of an alternative one at a given significance.~This assessment 
is crucial, as it contributes determining which experiments to prioritise in the coming years.~One important component of this assessment is to clarify quantitatively whether the simultaneous measurement of nuclear recoil energy spectrum and associated time distribution of the observed recoils is advantageous for a given experimental setup, or it is preferable to focus on the measurement of the energy spectrum alone.~This question must be addressed proactively, as building ultra-low background experiments that are stable over the time scale of one year (the period of the Earth revolution around the Sun) is a difficult experimental approach posing highly non trivial challenges in particular for cryogenic detectors.

The impact of timing information on the ability of direct detection experiments to distinguish DM models with nearly degenerate recoil spectra has been investigated for single-target experiments employing fluorine, germanium or xenon targets and with a relatively high energy threshold in~\cite{Witte:2016ydc}.~Using a Bayesian approach to model selection, Ref.~\cite{Witte:2016ydc} finds that including timing information may enhance this ability, but only with exposures beyond the expectation of the experiments currently running (indicated as Generation 2 experiment in ~\cite{Witte:2016ydc}).~The results of Ref.~\cite{Witte:2016ydc} apply to weakly interacting massive particles of mass around 50 GeV and cannot a priori be extended to DM at the GeV scale.~This extension would require lowering the assumed energy threshold below about 0.3 keV.

Similarly, the impact of timing information on the ability of direct detection experiments to identify a DM signal over an experimental background has not been investigated for low-threshold experiments.~Consequently, it remains unclear  whether it is worth taking the challenge of operating cryogenic experiments under stable conditions over a period of one year or longer.~Importantly, this is a timely question, as the low-threshold experiment CRESST has recently reported unexplained low-energy events in one of the detector modules~\cite{Abdelhameed:2019hmk}.~These start at the detector energy threshold, 30.1 eV, and extend up to 200 eV, inside the acceptance region in the light yield - recoil energy plane.~The CRESST collaboration reports that the excess is present in all the operated modules and that its energy distribution  varies from module to module, observation which disfavours a single common origin of this effect.~However, this excess limits CRESST sensitivity.~Consequently, it is reasonable to ask whether a time-dependent analysis improves the significance for signal identification and model selection in the presence of such background.

In this work, we investigate the impact of timing information on the potential for DM signal discovery and model selection of a low-threshold experiment.~The analysis is divided into two parts.~In the first one, we test the impact of timing information on the ability of a low-threshold experiment to discover a DM signal over an experimental background that resembles the population of low-energy events found in~\cite{Abdelhameed:2019hmk}.~The second part of the analysis focuses on DM model selection.~Here, we test the ability of a low-threshold experiment to reject a model where DM couples to nucleons via spin-independent interactions in favour of an alternative hypothesis where DM couples to nuclei via magnetic dipole interactions.~Our choice of alternative hypothesis is motivated by the fact that the time of maximum DM-induced nuclear recoil rate as a function of the minimum velocity required to transfer a given momentum in the scattering can be target-dependent in the case of magnetic dipole interactions, whereas it does not depend on the target for spin-independent interactions~\cite{DelNobile:2015tza}.~Consequently, a multi-target experiment like CRESST should be able to exploit this feature when comparing different DM models. ~Remarkably, we find that timing information has not a significant impact on DM signal discovery, but on model selection for experimental exposures that are within reach of next-generation, low-threshold direct detection experiments.~For example, we find that with an exposure of 460 g$\times$year the $p$-value for rejecting spin-independent interactions in favour of a magnetic dipole coupling is around 0.11, while it is about 0.06 when the time distribution of the event rate is available.~At the same time, our results show that a 90\% C.L. rejection of spin-independent interactions is within reach of a future upgrade of the CRESST experiment~\cite{CRESST:2015djg}.

This work is organised as follows.~In Sec.~\ref{sec:theory} we outline the theoretical framework used in our analysis.~In Sec.~\ref{sec:timing}, we critically review the timing information that is available to a low-threshold, multi-target experiment like CRESST.~In Sec.~\ref{sec:stat}, we introduce the statistical framework that we apply to assess the prospects for DM signal discover and model selection with a low-threshold experiment.~We present our results on DM signal discovery in Sec.~\ref{sec:signal} and on DM model selection in Sec.~\ref{sec:model}.~We finally summarise and conclude in Sec.~\ref{sec:conclusion}.~In an Appendix, we list useful expressions to compute scattering rates and cross sections in the case of spin 1/2 DM.

\section{Theoretical framework}
\label{sec:theory}
The differential rate of DM-nucleus scattering events per unit detector mass in a direct detection experiment can be expressed as an integral over DM particle velocities in the detector rest frame,
\begin{align}
\frac{{\rm d} R}{{\rm d} E_R} = \sum_T \xi_T \frac{\rho_\chi}{m_\chi m_T} \int_{|\vec v| \ge v_{\rm min}} {\rm d}^3 v \, |\vec v| f(\vec v,t) \, \frac{{\rm d}\sigma_T}{{\rm d} E_R} \,,
\label{eq:rate}
\end{align}
where $v_{\rm min}=\sqrt{2 m_T E_R}/(2 \mu_T)$ is the minimum velocity a DM particle must have to deposit an energy $E_R$ in the detector, $m_T$ and $\mu_T$ are the target nucleus and DM-nucleus reduced mass, respectively, $m_\chi$ is the DM particle mass and $\rho_\chi$ is the local DM density.~In multi-target detectors, like CRESST, the contribution to Eq.~(\ref{eq:rate}) of each element in the detector material is weighted by the corresponding mass fraction, $\xi_T$.~The differential rate of DM-nucleus scattering events also depends on the differential cross section for DM-target nucleus scattering, ${\rm d}\sigma_T/{\rm d} E_R$, and on the local DM velocity distribution expressed as a function of the DM velocity in the detector reference frame, $f(\vec{v},t)$.~The velocity distribution $f(\vec{v},t)$ is a periodic function of time, $t$, with period of one year.~As a result, the expected rate of nuclear recoils is a function of time with the same periodicity.~We model $f(\vec{v},t)$ by taking  the gravitational focusing of the Sun into account, which implies~\cite{Alenazi:2006wu},
\begin{equation}
f(\vec{v},t) = \tilde{f}(\vec{v_{\odot}}+\vec{v_{\infty}}[\vec{v}+\vec{V}_\oplus(t)])\,,
\end{equation}
where $\tilde{f}$ is the DM velocity distribution in the Galactic reference frame (where, by construction, the mean velocity of DM particles is zero), $\vec{v_\odot}\simeq (11,232,7)$~km~s$^{-1}$ is the relative velocity of the Sun with respect to the Galactic centre in Galactic coordinates, $\vec{v}$ is the DM particle velocity in the detector rest frame and $\vec{V}_{\oplus}(t)$ is the relative velocity of the Earth with respect to the Sun.~Consistently, we denote by $\vec{v_s}=\vec{v}+\vec{V}_\oplus$ the DM particle velocity in the Sun rest frame.~With this notation, the solar frame velocity $\vec{v_\infty}$ a DM particle must have at infinity to move with velocity $\vec{v_s}$ with respect to the Sun at the detector position is given by~\cite{Alenazi:2006wu},
\begin{equation}
\vec{v_\infty}[\vec{v_s}]=\frac{v_\infty^2 \vec{v_s} + v_\infty (G M_\odot/r_s)\vec{\hat{r}_s}-v_\infty\vec{v_s}
(\vec{v_s}\cdot \vec{\hat{r}_s})}{v_\infty^2+(G M_\odot/r_s)-v_\infty(\vec{v_s}\cdot \vec{\hat{r}_s})}\,,
\end{equation}
where $r_s$ is Earth-Sun distance, $\vec{\hat{r}_s}$ is the time dependent unit vector that points from the Sun to the Earth, $v_\infty^2=v^2-2GM_\odot/r_s$, $M_\odot$ is the mass of the Sun, and $G$ is Newton's constant.~Explicit expressions for $\vec{V_{\oplus}}$ and $\vec{\hat{r}_s}$ in Galactic coordinates can be found in \cite{Lee:2013wza} and \cite{McCabe:2013kea,Lee:2013xxa}, respectively.~For $\tilde{f}$, we assume a Gaussian distribution truncated at the escape velocity $v_{\rm esc},$
\begin{align}
\tilde{f}(\vec{\tilde{v}})=\frac{1}{N_{\rm esc}(2\pi \sigma_v^2)^{3/2}}\,\exp\left(-\frac{|\vec{\tilde{v}}|^2}{2 \sigma_v^2}\right)\Theta\left(v_{\rm esc}-|\vec{\tilde{v}}|\right)\,,
\label{eq:MB}
\end{align}
where $\vec{\tilde{v}}=\vec{v_{\odot}}+\vec{v_{\infty}}$.~Here, we assume the one-dimensional velocity dispersion $\sigma_v=164$~km~s$^{-1}$ and a local escape velocity of $v_{\rm esc},=544$~km~s$^{-1}$.~The normalisation factor, $N_{\rm esc}$, in Eq.~(\ref{eq:MB}) is given by,
\begin{equation}
N_{\rm esc}= \erf\left(\frac{v_{\rm esc}}{\sqrt{2} \sigma_v}\right) - \sqrt{\frac{2}{\pi}} \frac{v_{\rm esc}}{\sigma_v} \exp\left(-\frac{v_{\rm esc}^2}{2\sigma_v^2} \right) \,.
\label{eq:Nesc}
\end{equation}
\begin{table}[t]
    \centering
    \begin{tabular*}{\columnwidth}{@{\extracolsep{\fill}}ll@{}}
    \toprule
      $\mathcal{O}_1 = \mathds{1}_{\chi N}$ & $\mathcal{O}_9 = i\mathbf{S}_\chi\cdot\left(\mathbf{S}_N\times\frac{ \mathbf{q}}{m_N}\right)$  \\
        $\mathcal{O}_3 = i\mathbf{S}_N\cdot\left(\frac{ \mathbf{q}}{m_N}\times \mathbf{v}^{\perp}\right)$ \hspace{2 cm} &   $\mathcal{O}_{10} = i\mathbf{S}_N\cdot\frac{ \mathbf{q}}{m_N}$   \\
        $\mathcal{O}_4 = \mathbf{S}_{\chi}\cdot \mathbf{S}_{N}$ &   $\mathcal{O}_{11} = i\mathbf{S}_\chi\cdot\frac{ \mathbf{q}}{m_N}$   \\                                                                             
        $\mathcal{O}_5 = i\mathbf{S}_\chi\cdot\left(\frac{ \mathbf{q}}{m_N}\times \mathbf{v}^{\perp}\right)$ &  $\mathcal{O}_{12} = \mathbf{S}_{\chi}\cdot \left(\mathbf{S}_{N} \times \mathbf{v}^{\perp} \right)$ \\                                                                                                                 
        $\mathcal{O}_6 = \left(\mathbf{S}_\chi\cdot\frac{ \mathbf{q}}{m_N}\right) \left(\mathbf{S}_N\cdot\frac{\hat{{\bf{q}}}}{m_N}\right)$ &  $\mathcal{O}_{13} =i \left(\mathbf{S}_{\chi}\cdot  \mathbf{v}^{\perp}\right)\left(\mathbf{S}_{N}\cdot \frac{ \mathbf{q}}{m_N}\right)$ \\   
        $\mathcal{O}_7 = \mathbf{S}_{N}\cdot  \mathbf{v}^{\perp}$ &  $\mathcal{O}_{14} = i\left(\mathbf{S}_{\chi}\cdot \frac{ \mathbf{q}}{m_N}\right)\left(\mathbf{S}_{N}\cdot  \mathbf{v}^{\perp}\right)$  \\
        $\mathcal{O}_8 = \mathbf{S}_{\chi}\cdot  \mathbf{v}^{\perp}$  & $\mathcal{O}_{15} = -\left(\mathbf{S}_{\chi}\cdot \frac{ \mathbf{q}}{m_N}\right)\left[ \left(\mathbf{S}_{N}\times  \mathbf{v}^{\perp} \right) \cdot \frac{ \mathbf{q}}{m_N}\right] $ \\       
    \bottomrule
    \end{tabular*}
    \caption{Interaction operators defining the non-relativistic effective theory of spin 1/2 DM-nucleon interactions~\cite{Fan:2010gt,Fitzpatrick:2012ix}.~$\mathbf{S}_N$ ($\mathbf{S}_\chi$) is the nucleon (DM) spin, $\mathbf{v}^\perp=\mathbf{v}-\mathbf{q}/(2\mu_N)$, where $\mu_N$ is the DM-nucleon reduced mass, is the transverse relative velocity and $\mathds{1}_{\chi N}$ is the identity in the DM-nucleon spin space.}
\label{tab:operators}
\end{table}The differential cross section for DM-nucleus scattering, ${\rm d}\sigma_T/{\rm d} E_R$ encodes all particle and nuclear physics inputs required to evaluate Eq.~(\ref{eq:rate}).
~Within the non relativistic effective theory of DM-nucleon interactions, ${\rm d}\sigma_T/{\rm d} E_R$ can be expressed as follows,
\begin{align}
\frac{{\rm d}\sigma_{T}}{{\rm d}E_R} = \frac{2m_T}{(2J_T+1)v^2} \sum_k\sum_{\tau,\tau'}\left(\frac{q^2}{m_N^2}\right)^{\ell_k} R^{\tau\tau'}_k\left(v_T^{\perp 2}, {q^2 \over m_N^2} \right) W_k^{\tau\tau'}(q^2)  \,,
\label{eq:dsigma} 
\end{align}
where $J_T$ is the target nucleus spin, $v=|\mathbf{v}|$, $v_T^{\perp 2}=v^2-q^2/(4\mu_T^2)$, $q=\sqrt{2 m_T E_R}$ is the momentum transfer and $m_N$ is the nucleon mass.~At most eight DM and nuclear response functions, $R^{\tau\tau'}_k$ and $W^{\tau\tau'}_k$, $k=M,\Sigma',\Sigma'',\Phi'', \Phi'' M, \tilde{\Phi}', \Delta, \Delta \Sigma'$, respectively, can appear in Eq.~(\ref{eq:dsigma})~\cite{Fitzpatrick:2012ix}.~The $R^{\tau\tau'}_k$ functions depend on $q^2/m_N^2$,  
$v_T^{\perp 2}$  and on the coupling constants for DM-nucleon interactions, $c_j^\tau$.~They are known analytically, and we list them for spin 1/2 DM in the Appendix.~Here, the index $j$ labels the type of DM-nucleon interaction~\cite{Fitzpatrick:2012ix}.~Assuming one-body DM-nucleon interactions, there are $4+20 J_\chi$ interaction types that are invariant under Galilean transformations and spatial rotations for a DM particle of spin $J_\chi$~\cite{Gondolo:2020wge}.~We list the 14 independent interaction operators arising for spin 1/2 DM in Tab.~\ref{tab:operators}.~The eight nuclear response functions $W_k^{\tau\tau'}$ in Eq.~(\ref{eq:dsigma}) are quadratic in reduced matrix elements of nuclear charges and currents, and must be computed numerically.~This calculation is performed within the nuclear shell model for heavy targets, such as xenon.~However, for light nuclei {\it ab initio} methods have recently been applied~\cite{Gazda:2016mrp}.~The indexes $\tau$ and $\tau'$ run from 0 to 1:~0 corresponds to ``isoscalar'' interactions and 1 to ``isovector'' interactions~\cite{Fitzpatrick:2012ix}.~In terms of, e.g.~$c_1^\tau$, the neutron to proton coupling ratio can be expressed as $r=c_p/c_n=(c_1^0-c_1^1)/(c_1^0+c_1^1)$.~Finally, $\ell_k=0$ for $k=M,\Sigma',\Sigma''$, and $\ell_k=1$ otherwise.

The way the cross section ${\rm d}\sigma_T/{\rm d} E_R$ depends on $v$ determines how the nuclear recoil rate depends on $t$.~Inspection of Eq.~(\ref{eq:R}) shows that there are four ways the velocity $v$ can enter the cross section ${\rm d}\sigma_T/{\rm d} E_R$.~The interactions labelled by $\mathcal{O}_1$, $\mathcal{O}_7$, $\mathcal{O}_8$ and $\mathcal{O}_{11}$ in the literature~\cite{Fitzpatrick:2012ix} and here associated with the coupling constants $c_1^\tau$, $c_7^\tau$, $c_8^\tau$ and $c_{11}^\tau$ are representative of these four classes.~The corresponding DM-nucleus scattering cross section reads as,
\begin{align}
\frac{{\rm d}\sigma_{T}}{{\rm d}E_R}\Bigg|_{\mathcal{O}_1}&= \frac{2 m_T}{(2J_T +1 ) v^2} \sum_{\tau \tau'} c_1^\tau c_1^{\tau'} W_{M}^{\tau\tau'}(q^2)\equiv \frac{a(q^2)}{v^2} \,,\nonumber \\
\frac{{\rm d}\sigma_{T}}{{\rm d}E_R}\Bigg|_{\mathcal{O}_7}&= \frac{2 m_T}{(2J_T +1 ) v^2} \sum_{\tau \tau'} c_7^\tau c_7^{\tau'} \frac{(v^2 - v_{\rm min}^2)}{8}W_{M}^{\tau\tau'}(q^2)  \equiv b(q^2)\left(1-\frac{v_{\rm min}^2}{v^2}\right)  \,, \nonumber\\
\frac{{\rm d}\sigma_{T}}{{\rm d}E_R}\Bigg|_{\mathcal{O}_8}&= \frac{2 m_T}{(2J_T +1 ) v^2} \sum_{\tau \tau'} \frac{c_8^\tau c_8^{\tau'}}{4} 
\left[v^2 W_{M}^{\tau\tau'}(q^2) - v_{\rm min}^2 \left(W_{M}^{\tau\tau'}(q^2) - 4\frac{\mu^2_T}{m^2_N} W_{\Delta}^{\tau\tau'}(q^2)\right) \right]\,, \nonumber\\
&\equiv c(q^2)\left(1-\frac{v_{\rm min}^2}{v^2}\right) + d(q^2) \frac{v_{\rm min}^2}{v^2} \nonumber\\
\frac{{\rm d}\sigma_{T}}{{\rm d}E_R}\Bigg|_{\mathcal{O}_{11}}&= \frac{2 m_T}{(2J_T +1 ) v^2} \sum_{\tau \tau'} \frac{c_{11}^\tau c_{11}^{\tau'}}{4}\frac{q^2}{m^2_N}W_{M}^{\tau\tau'}(q^2) \equiv e(q^2) \frac{v_{\rm min}^2}{v^2}\,,
\label{eq:eft}
\end{align}
where we assumed that the DM particle has spin $J_\chi=1/2$.~In Eq.~(\ref{eq:eft}), we also emphasised the dependence on $v$ by introducing the functions $a$, $b$, $c$, $d$, and $e$ which are implicitly defined via the above equation.~We refer to~\cite{Fan:2010gt,Fitzpatrick:2012ix} for further details on the effective theory of DM-nucleon interactions and to~\cite{Hoferichter:2015ipa,Bishara:2016hek} for a discussion on its limitations and extensions.~Here, we just mention that the $\mathcal{O}_1$ interaction is the familiar spin-independent interaction~\cite{Fitzpatrick:2012ix} and that the interactions $\mathcal{O}_7$, $\mathcal{O}_8$ and $\mathcal{O}_{11}$ arise from the non-relativistic reduction of simplified models.~They can be the leading DM-nucleon interactions in specific ranges for the DM mass~\cite{Catena:2017xqq,Baum:2017kfa,Baum:2018lua,Baum:2018sxd}.

While in the non-relativistic effective theory of DM-nucleon interactions the coupling constants $c_j^\tau$ are considered as independent, in concrete models one generically expects to generate specific combinations of operators in Tab.~\ref{tab:operators}.~In particular, this applies to the DM magnetic dipole model.~The model assumes that DM is made of spin 1/2 particles and is characterised by the interaction Lagrangian $\mathscr{L} = \frac{1}{2}\lambda_\chi \bar{\chi} \sigma^{\mu \nu} \chi F_{\mu\nu}$, where $\chi$ is the DM spinor and $F_{\mu\nu}$ the electromagnetic field strength tensor.~Here, the coupling constant $\lambda_\chi$ has dimension GeV$^{-1}$.~From this Lagrangian one finds the non-relativistic amplitude for DM-nucleon scattering~\cite{Kavanagh:2018xeh},\begin{align}
\mathcal{M}&= e \lambda_\chi  \left[2 m_N Q_N \langle \mathcal{O}_1\rangle + 4 m_\chi g_N \langle \mathcal{O}_4\rangle + \frac{8 m^2_N m_\chi Q_N}{|\mathbf{q}|^2} \langle \mathcal{O}_5\rangle -  \frac{4 m^2_N m_\chi g_N}{|\mathbf{q}|^2}  \langle \mathcal{O}_6\rangle
    \right]\,,
    \label{eq:M}
\end{align}
where $Q_N$, $N=p,n$, is the nucleon electric charge (0 for neutrons and 1 for protons), $g_N$, $N=p,n$, is the nucleon $g$-factor, and angle brackets denote matrix elements between two-component nucleon and DM spinors.~Eq.~(\ref{eq:M}) shows that the amplitude for DM-nucleon scattering in the DM magnetic dipole model is a linear combination of matrix elements of the operators in Tab.~\ref{tab:operators} where some of the coefficients are constant while others scale like $1/|\mathbf{q}|^2$ with the momentum transfer.~Consequently, if one promotes the coupling constants $c_j^\tau$ to functions of the momentum transfer, the DM magnetic dipole model arises from a specific combination of operators in Tab.~\ref{tab:operators}.~The associated cross section for DM-nucleus scattering reads as follows,
\begin{align}
\frac{{\rm d}\sigma_{T}}{{\rm d}E_R} &= \frac{8 m_T}{(2J_T +1 ) v^2}  \alpha \pi \lambda_\chi^2 \Bigg[ \Bigg( \frac{1}{m_\chi^2} - \frac{1}{\mu_T^2} + \frac{1}{\mu_T^2} \frac{v^2}{v^2_{\rm min}} \Bigg) W_{M}^{pp} + \frac{1}{m_N^2}  \Bigg( \tilde{\mu}_p^2 W_{\Sigma'}^{pp} +2 \tilde{\mu}_p \tilde{\mu}_n W_{\Sigma'}^{np} \nonumber\\ &+ \tilde{\mu}_n^2  W_{\Sigma'}^{nn}\Bigg) +4 W_{\Delta}^{pp} -4\tilde{\mu}_pW_{\Delta \Sigma'}^{pp} - 4\tilde{\mu}_n W_{\Delta \Sigma'}^{pn}\Bigg] \,,
\label{eq:mddm}
\end{align}
where $\alpha$ is the fine structure constant, while $\tilde{\mu}_N=g_N/2$, with $\tilde{\mu}_p=2.8$ and $\tilde{\mu}_n=-1.9$, is the dimensionless magnetic moment of the nucleon~\cite{Gresham:2014vja}.~The nuclear response functions in Eq.~(\ref{eq:mddm}) are related to the ones in Eq.~(\ref{eq:eft}) by $W_{M}^{pp}=W_{M}^{00}+2W_M^{01}+W_{M}^{11}$ and analogous expressions~\cite{Fitzpatrick:2012ix}.

We conclude this section by specifying our choices of nuclear targets and response functions.~So far, the CRESST experiment has exploited CaWO$_4$, Li$_2$MoO$_4$ and Al$_2$O$_3$ targets.~Here, we primarily focus on detectors employing Al$_2$O$_3$ crystals.~This choice is motivated by the fact that Al$_2$O$_3$ is made of two light elements, aluminium and oxygen, and  Al$_2$O$_3$ detectors can operate with  energy thresholds as low as 19 eV~\cite{Angloher:2017sxg}.~Both properties make Al$_2$O$_3$ a promising target to search for light DM particles of mass of a few GeV.~Furthermore, aluminium has spin 5/2 and the corresponding $W_{\Sigma'}^{\tau\tau'}$, $W_{\Delta}^{\tau\tau'}$ and $W_{\Delta \Sigma'}^{\tau\tau'}$ functions are different from zero.~Consequently, Al$_2$O$_3$ detectors can be used to probe a wide range of DM-nucleon interactions.~In the calculations reported in Sec.~\ref{sec:amplitude}, Sec.~\ref{sec:harmonics} and Sec.~\ref{sec:target}, as well as in Secs.~\ref{sec:signal} and \ref{sec:model}, we use the nuclear form factors of~\cite{Catena:2015uha} for aluminium and oxygen (as well as calcium in some investigations).~For the same elements, we use Helm form factors when treating the background amplitude as a nuisance parameter (see Sec.~\ref{sec:stat}), as they are faster to evaluate.~Results obtained by using Helm form factors lead to overestimate the expected number of DM-induced nuclear recoil events by about 10\%.~Finally, when comparing the timing information available to a Al$_2$O$_3$ multi-target experiment with the one accessible with a fluorine, germanium, iodine, or xenon detector, for the latter we use the nuclear form factors provided by the \texttt{DMFormFactor} code~\cite{Anand:2013yka}.

\section{Timing information in multi-target detectors}
\label{sec:timing}
In this section, we critically review the timing information that is available to low-threshold multi-target DM direct detection experiments when DM couples to nuclei via one of the interactions in Tab.~\ref{tab:operators}.~This timing information is contained in the coefficients of the Fourier series expansion of the differential rate of nuclear recoil events in Eq.~(\ref{eq:rate}).~The time-averaged rate, $A_0$, and the annual modulation amplitude, $A_1$, are the first two coefficients in this expansion.~Starting from this observation, in Sec.~\ref{sec:amplitude} we characterise the interactions in Tab.~\ref{tab:operators} in terms of the predicted $A_1(v_{\rm min})$ curve.~In~Sec.~\ref{sec:harmonics}, we show that most of the timing information is contained in $A_0$ and $A_1$ for all interactions in Tab.~\ref{tab:operators}, and hence higher-order harmonics can be neglected.~Finally, in Sec.~\ref{sec:target}, we characterise the interactions in Tab.~\ref{tab:operators} in terms of the predicted $t_{\rm max}(v_{\rm min})$ curve introduced in~\cite{DelNobile:2015tza}, where $t_{\rm max}$ is the time of maximum nuclear recoil rate.~At the end of Sec.~\ref{sec:target}, we also comment on the expected amplitude vs $v_{\rm min}$ and $t_{\rm max}$ vs $v_{\rm min}$ curves in the case of DM-nucleus magnetic dipole interactions.~The timing information that we review in this section will then be used in the following two sections to assess the prospects for DM signal discovery and model selection with next-generation low-threshold detectors.

\subsection{Annual modulation amplitude}
\label{sec:amplitude}
Since the vectors $\vec{V_{\oplus}}$ and $\vec{\hat{r}_s}$ are periodic functions of $t$ with period of one year, the rate of DM-nucleus scattering events, Eq.~(\ref{eq:rate}), can be expanded in Fourier series around a reference time $t_0>0$,
\begin{equation}
\frac{{\rm d} R}{{\rm d} E_R} = A_0 + \sum_{n=1}^{\infty} A_n \cos n\omega(t-t_0) + \sum_{n=1}^{\infty} B_n \sin n\omega(t-t_0)\,,
\label{eq:fourier}
\end{equation}
where $\omega=2\pi/\textrm{year}$ and for $t_0$ we choose the value of $t$ that maximises the vector $\vec{v_{\odot}}+\vec{V_{\oplus}}$.~The expansion coefficients $A_0$, $A_n$ and $B_n$ depend on the nuclear recoil energy or, equivalently, on $v_{\rm min}$.~They also depend on the differential cross section for DM-nucleus scattering, which in turn depends on how DM couples to nuclei.~If the DM velocity distribution in the Galactic reference frame is isotropic, then $B_n=0$~\cite{Freese:2012xd}.~Furthermore, when $f(\vec{v},t)$ changes slowly over velocity variations of the order of $|\vec{V}_\oplus|$, then $A_0\gg A_1\gg A_{n\ge 2}$~\cite{Lee:2013xxa}.~The existence of this hierarchy is well-known in the case of the familiar spin-independent and spin-dependent interactions.~In Sec.~\ref{sec:harmonics}, we show that it also applies to all DM-nucleon interactions considered in Tab.~\ref{tab:operators}.~In the ``single cosine approximation'', all expansion coefficients in Eq.~(\ref{eq:fourier}) but $A_0$ and $A_1$ are set to zero.~Within this approximation, the modulation amplitude, $A_1$, can be written as
\begin{align}
A_1(v_{\rm min}) = \frac{1}{2} \left[ \frac{{\rm d} R}{{\rm d} E_R} \left( v_{\rm min}, t_0 \right) -  \frac{{\rm d} R}{{\rm d} E_R} \left( v_{\rm min}, t_0+\mbox{year}/2 \right) \right]\,,
\label{eq:A1}
\end{align}
where we emphasised the dependence of the differential rate ${\rm d} R/{\rm d} E_R$ on $v_{\rm min}$ and the time of the year.

\begin{figure}[t]
\begin{center}
\includegraphics[width=\textwidth]{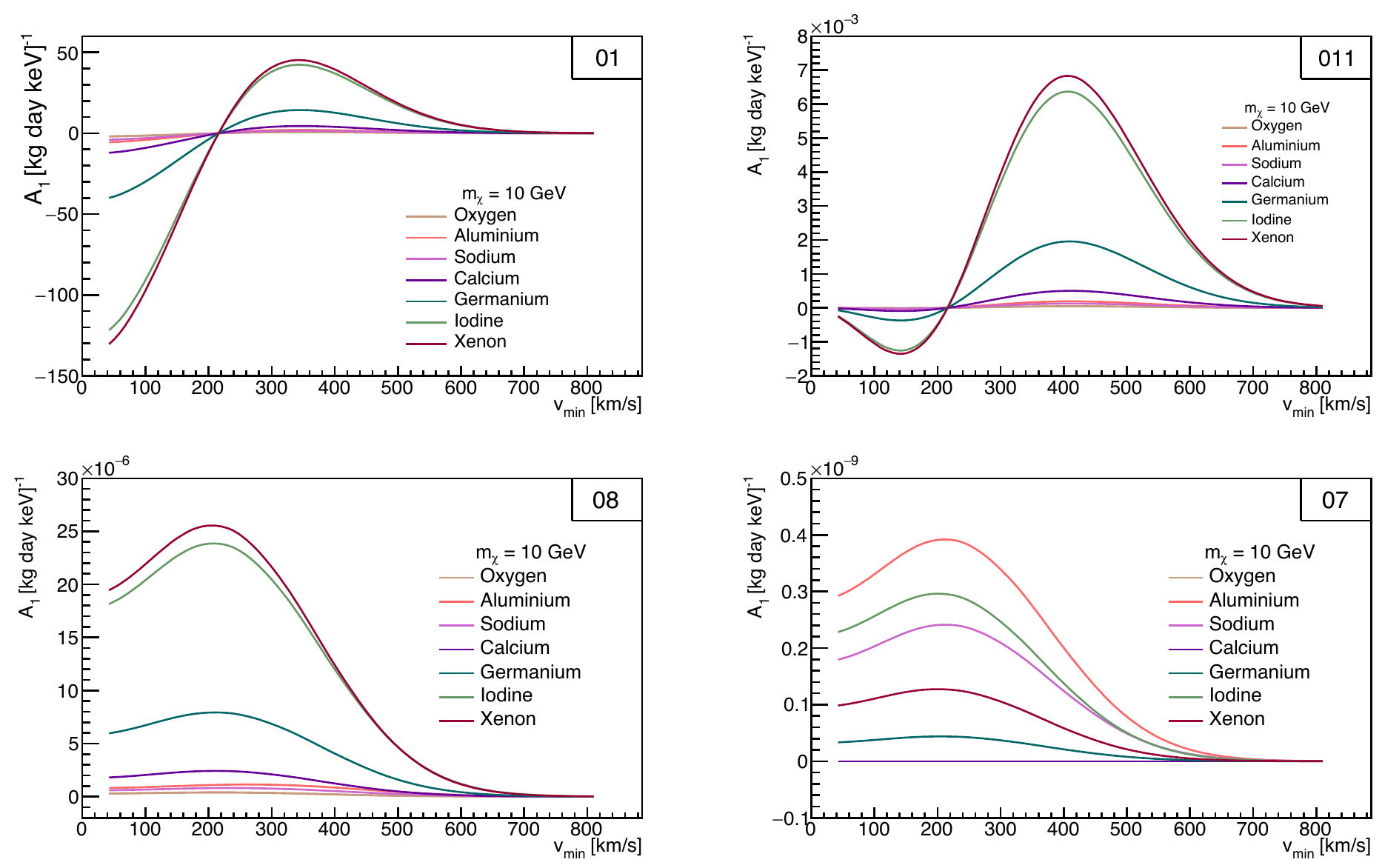}
\end{center}
\caption{Modulation amplitude $A_1$ as a function of the minimum velocity $v_{\rm min}$ for $m_\chi$=10~GeV and the four interactions $\mathcal{O}_1$, $\mathcal{O}_7$, $\mathcal{O}_8$ and $\mathcal{O}_{11}$.~Different colours refers to distinct targets.~For the interactions $\mathcal{O}_1$ and $\mathcal{O}_{11}$ (top panels), $A_1$ changes sign for $v_{\rm min}$ around 200 km~s$^{-1}$.~For the interactions $\mathcal{O}_7$ and $\mathcal{O}_{8}$ (bottom panels) $A_1>0$ for every $v_{\rm min}$.~Here, we assume a positive reference time $t_0$ (see Eq.~(\ref{eq:A1})).}  
\label{fig:A1}
\end{figure}
\begin{figure}[t]
\begin{center}
\includegraphics[width=\textwidth]{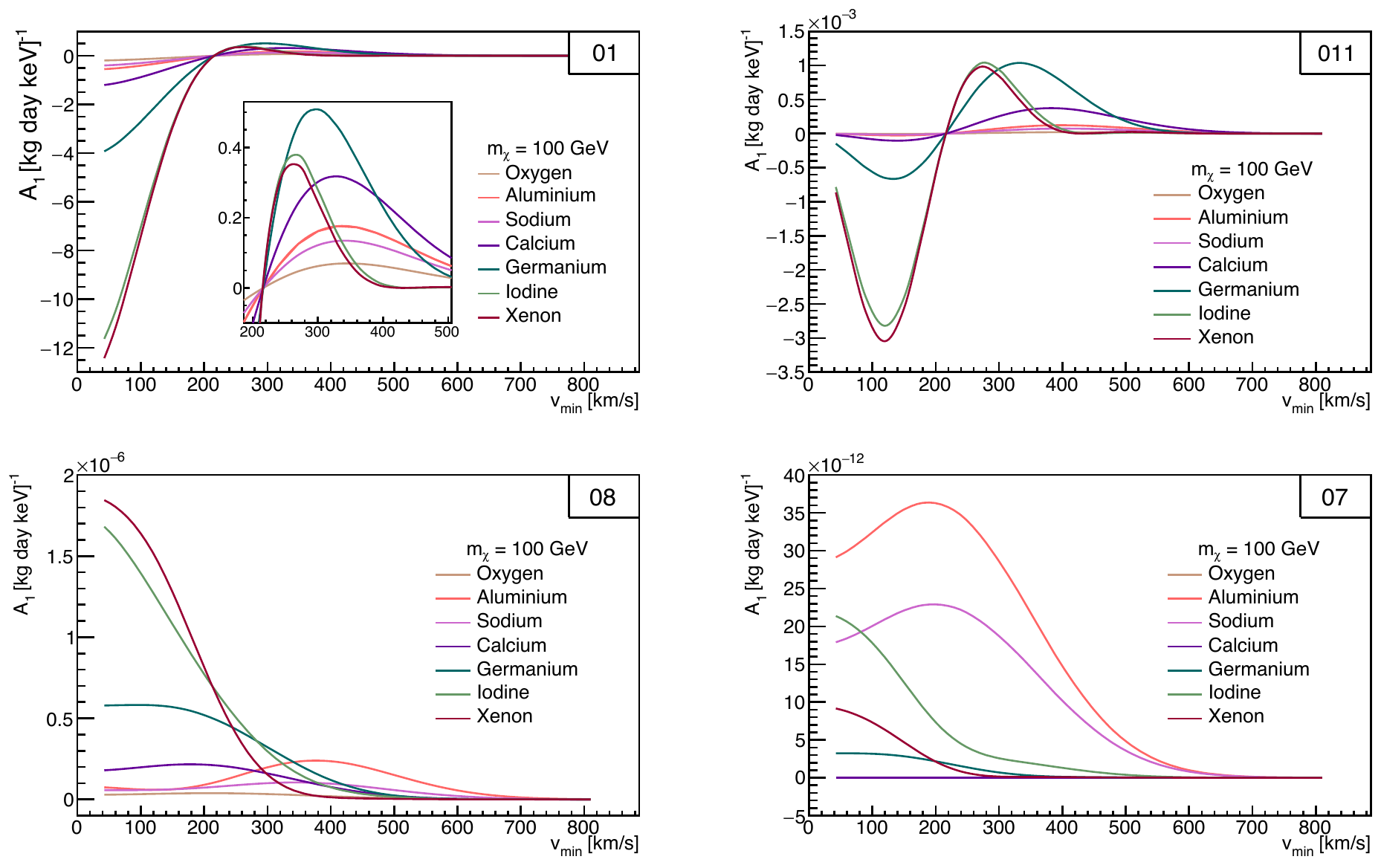}
\end{center}
\caption{Same as Fig.~\ref{fig:A1}, but now for $m_\chi=100$~GeV.}  
\label{fig:A1_2}
\end{figure}
We now characterise the interactions in Tab.~\ref{tab:operators} in terms of the predicted $A_1(v_{\rm min})$ curve.~Since the dependence on $v$ of the DM-nucleus scattering cross section determines the time-dependence of the nuclear recoil rate in Eq.~(\ref{eq:rate}), we can restrict our analysis to the four representative interactions in Eq.~(\ref{eq:eft}).~For the latter, the four panels in Fig.~\ref{fig:A1} show $A_1$ as a function of $v_{\rm min}$.~We set the DM mass to $m_\chi=10$~GeV and one coupling constant at the time to $1/m_V^2$, where $m_V=246$~GeV.~Each panel contains seven lines of different colours corresponding to distinct target materials:~oxygen, aluminium, sodium, calcium, germanium, iodine and xenon.~In our calculations, we take the isotopic abundance of the different elements into account.~For the interactions $\mathcal{O}_1$ and $\mathcal{O}_{11}$ (top panels), the modulation amplitude changes sign for $v_{\rm min}$ around 200 km~s$^{-1}$.~This is consistent with having assumed $t_0>0$ and $A_1$ unconstrained.~On the other hand, we could have assumed $A_1>0$ and $t_0$ unconstrained.~This second convention would have implied a change of sign for $t_0$, i.e.~an ``inversion of phase'', for $v_{\rm min}\sim$~200 km~s$^{-1}$.~For the interactions $\mathcal{O}_7$ and $\mathcal{O}_{8}$ (bottom panels) there is no inversion of phase, i.e.~$A_1>0$ for every $v_{\rm min}$.~Fig.~\ref{fig:A1} also shows that for the $\mathcal{O}_1$, $\mathcal{O}_8$ and $\mathcal{O}_{11}$ interactions, the heavier the target the larger $A_1$.~This is expected, as the corresponding scattering cross section depends on $W_{M}^{\tau \tau'}$, and, therefore, scales with the number of nucleons squared in the small momentum transfer limit.~In contrast, for the spin-dependent $\mathcal{O}_7$ interaction, aluminium exhibits the largest modulation amplitude.~Fig.~\ref{fig:A1_2} shows $A_1$ as a function of $v_{\rm min}$ for the same interactions and targets as Fig.~\ref{fig:A1}, but now for $m_\chi=100$~GeV.~Changing the DM particle mass, our results remain qualitatively the same.~Both Fig.~\ref{fig:A1} and Fig.~\ref{fig:A1_2} assume an exposure of 1~kg$\times$day.

\begin{figure}[t]
\begin{center}
\includegraphics[width=\textwidth]{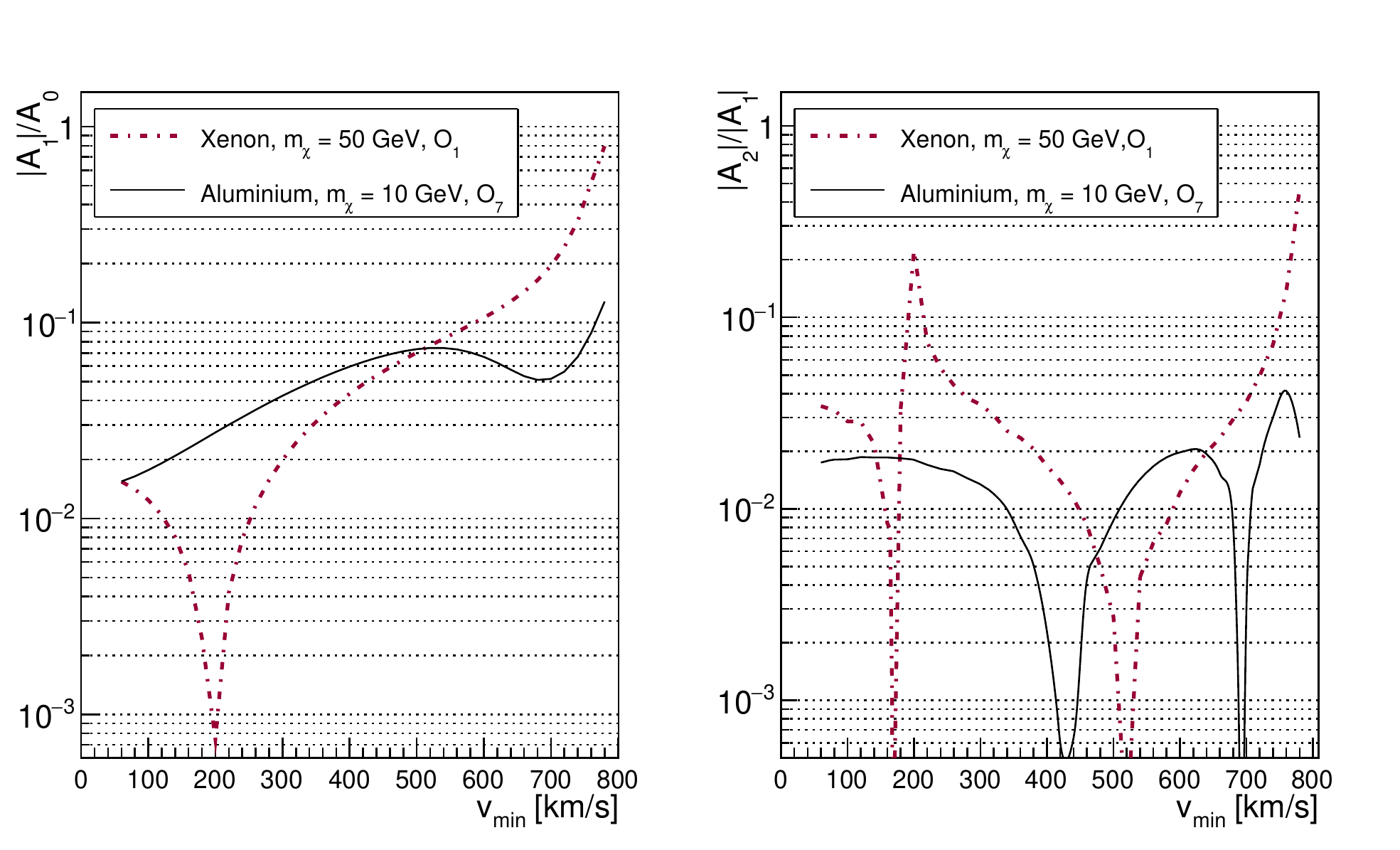}
\end{center}
\caption{Ratios $|A_1|/A_0$ (left panel) and $|A_2|/|A_1|$ (right panel) as a function of $v_{\rm min}$.~Red dot-dashed lines correspond to a DM particle of mass $50$~GeV scattering on a xenon target via $\mathcal{O}_1$.~Black solid lines refers to a DM particle of mass 10 GeV scattering on an aluminium target via $\mathcal{O}_7$.~The hierarchy $A_0\gg A_1\gg A_{2}$ is valid at all $v_{\rm min}$ for interactions where $A_1$ does not change sign, like $\mathcal{O}_7$.~It is valid at all $v_{\rm min}$ but around 200 km~s$^{-1}$ for interactions where $A_1$ changes sign around 200 km~s$^{-1}$, like $\mathcal{O}_1$.}  
\label{fig:harmonics}
\end{figure}
\subsection{Higher-order harmonics}
\label{sec:harmonics}
As we have seen in the previous subsection, the effective theory of DM-nucleon interactions predicts two families of modulation amplitude.~In the first one, $A_1$ changes sign as a function of $v_{\rm min}$.~In the second one, $A_1>0$ for every $v_{\rm min}$.~In this subsection we investigate the validity of the single cosine approximation, i.e.~the hierarchy $A_0\gg A_1\gg A_{n\ge 2}$, for the two families of interactions.~We focus on $\mathcal{O}_1$ and $\mathcal{O}_{7}$ as representatives of the first and second family, respectively.

Fig.~\ref{fig:harmonics}, left panel (right panel), shows the ratio $|A_1|/A_0$ ($|A_2|/|A_1|$) as a function of $v_{\rm min}$.~Here, we consider two benchmark scenarios:~1) A DM particle with interactions of $\mathcal{O}_1$ type and mass of $50$~GeV scattering on a xenon target (red, dot-dashed line).~2) A DM particle of mass 10 GeV interacting via the $\mathcal{O}_7$ interaction with an aluminium target (black, solid line).~The red lines in Fig.~\ref{fig:harmonics} reproduce the results of~\cite{Lee:2013xxa}, which we use to validate our calculations.~The black lines correspond to results obtained here and of interest for a low-threshold detector operating Al$_2$O$_3$ crystals.~The spike around 200 km~s$^{-1}$ in the $|A_1|/A_0$ curve for $\mathcal{O}_1$ reflects the change of sign of $A_1$ for this interaction.~Similarly, the smooth $|A_1|/A_0$ curve associated with $\mathcal{O}_7$ follows from the constant sign of $A_1$ in this case.~As far as the $|A_2|/|A_1|$ curve is concerned, we find two zeros at, respectively, 160-180~km~s$^{-1}$ and 500-540~km~s$^{-1}$ for the $\mathcal{O}_1$ interaction, and at 415-420~km~s$^{-1}$ and 690-700~km~s$^{-1}$ for the $\mathcal{O}_7$ interaction.~Consequently, we conclude that the single cosine approximation is valid at all $v_{\rm min}$ for interactions where $A_1$ does not change sign, like $\mathcal{O}_7$.~We also find that for interactions like $\mathcal{O}_1$  the single cosine approximation is valid at all $v_{\rm min}$ values but around 200 km~s$^{-1}$, as expected.~While we present our results focusing on two specific combinations of model parameters, it is important to stress that at a given $v_{\rm min}$, and for single interactions, the ratios $|A_1|/A_0$ and $|A_2|/|A_1|$ do not depend on DM particle and target mass, as these are either reabsorbed in the definition of $v_{\rm min}$ or factored out in time-independent pre-factors; see Eq.~(\ref{eq:rate2}) below\footnote{This is not exactly true for the category $\mathcal{O}_8$ ($\mathcal{O}_5$ ), since $\eta$ and $\tilde{\eta}$ present different target dependent coefficients. However, Fig.~\ref{fig:tmax} shows that the term proportional to $\tilde{\eta}$ dominates, therefore treating the ratios between higher order harmonics as independent from model parameters is a good approximation.}.

\subsection{Target dependence of the time of maximum rate}
\label{sec:target}
We now characterise the interactions in Tab.~\ref{tab:operators} in terms of the predicted $t_{\rm max}(v_{\rm min})$ curve, extending previous results to target nuclei of interest for the CRESST experiments.~We start by noticing that to evaluate the rate in Eq.~(\ref{eq:rate}) with differential cross sections for DM-nucleus scattering given in Eqs.~(\ref{eq:eft}) and (\ref{eq:mddm}), we have to compute one or two of the integrals, 
\begin{align}
\eta(v_{\rm min},t) &= \int_{v \ge v_{\rm min}} {\rm d}^3 v \,\frac{f(\vec v,t)}{v} \,,\\
\tilde{\eta}(v_{\rm min},t) &= \int_{v \ge v_{\rm min}} {\rm d}^3 v \,v f(\vec v,t) \,.
\end{align}
In terms of the $\eta(v_{\rm min},t)$ and $\tilde{\eta}(v_{\rm min},t)$ functions, the expected rate of nuclear recoils can be written as
\begin{align}
\frac{{\rm d} R}{{\rm d} E_R} = \mathscr{A}(v_{\rm min},m_T) \, \eta(v_{\rm min},t)+ \mathscr{B}(v_{\rm min},m_T) \,\tilde{\eta}(v_{\rm min},t) \,.
\label{eq:rate2}
\end{align}
When $\mathscr{A}(v_{\rm min},m_T)\neq0$ and $\mathscr{B}(v_{\rm min},m_T)=0$, as in the case of the familiar spin-independent interaction, the time of maximum rate, $t_{\rm max}$, is entirely determined by $\eta(v_{\rm min},t)$ and, as such, it is a universal, i.e.~target-independent function of $v_{\rm min}$.~The same applies to models where $\mathscr{B}(v_{\rm min},m_T)\neq0$ and $\mathscr{A}(v_{\rm min},m_T)=0$.~However, if both $\mathscr{A}(v_{\rm min},m_T)$ and $\mathscr{B}(v_{\rm min},m_T)$ are different from zero and depend on $m_T$, and if the two terms in Eq.~(\ref{eq:rate2}) have comparable size, then the function $t_{\rm max}(v_{\rm min})$ can depend on the target the DM particle scatters on.~In this latter case, there might be a range of $v_{\rm min}$ where the rate of DM-induced nuclear recoils observed by a multi-target detector depends on time via the superposition of two annual modulations of different phase.~Therefore, it is important to understand whether the target dependence of $t_{\rm max}(v_{\rm min})$ is a generic feature of DM-nucleus interactions, or it can only occur in specific models.~Here, we address this question within the effective theory of DM-nucleon interactions.~First, we compute $t_{\rm max}(v_{\rm min})$ for single interactions in Tab.~\ref{tab:operators}.~Then we focus on the specific linear combination of interaction operators corresponding to the DM magnetic dipole interaction model.~This latter calculation extends the results of Ref.~\cite{DelNobile:2015tza} to aluminium, calcium and oxygen.
\begin{figure}[t]
\begin{center}
\includegraphics[width=0.9\textwidth]{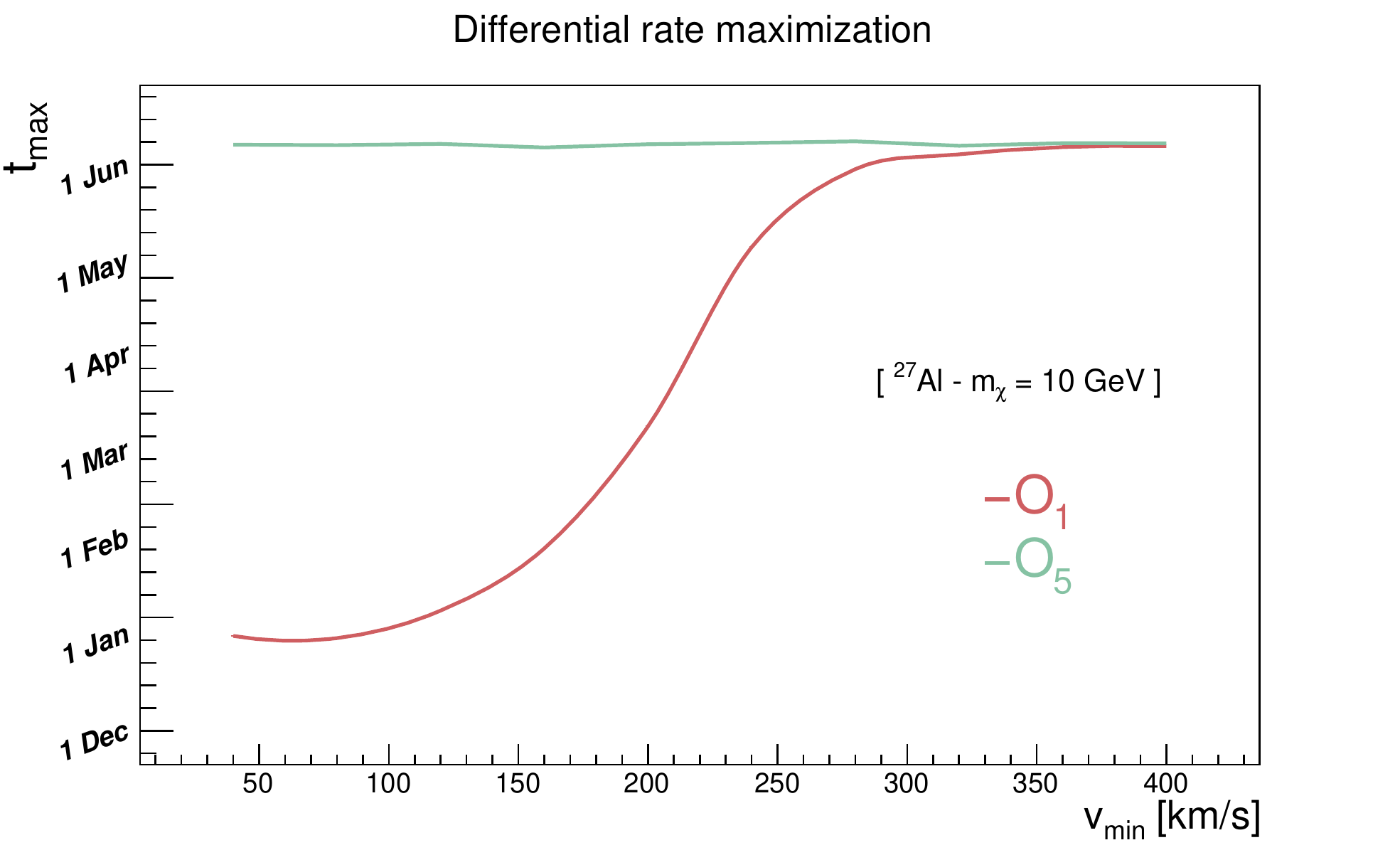}
\end{center}
\caption{Time of maximum, $t_{\rm max}$, for the differential rate in Eq.~(\ref{eq:rate}) as a function of $v_{\rm min}$ for the interactions $\mathcal{O}_1$ and $\mathcal{O}_5$.~In the former case Eq.~(\ref{eq:rate}) is proportional to the $\eta$ function, in the latter to the $\tilde{\eta}$ function.~In both cases, we assume aluminium as a target material and a DM particle mass of 10~GeV.}  
\label{fig:tmax}
\end{figure}

Inspection of Eq.~(\ref{eq:eft}) and Eq.~(\ref{eq:R}) shows that $\mathscr{A}(v_{\rm min},m_T)$ and $\mathscr{B}(v_{\rm min},m_T)$ are simultaneously different from zero for the $\mathcal{O}_5$ and $\mathcal{O}_8$ interactions only.~At the same time, by explicitly evaluating the $\mathscr{A}(v_{\rm min},m_T)$ and $\mathscr{B}(v_{\rm min},m_T)$ functions we find that in both cases the first term in Eq.~(\ref{eq:rate2}) always dominates over the second one.
~As a result, in models where DM couples to nuclei via one of the operators in Tab.~\ref{tab:operators} at the time, $t_{\rm max}(v_{\rm min})$ is a universal, target-independent curve.~We also find that all operators with differential cross section scaling like $a(q^2)/v^2$ and $e(q^2) v_{\rm min}^2/v^2$ predict a $t_{\rm max}(v_{\rm min})$ curve modulating from about January 1st at small $v_{\rm min}$ (when gravitational focusing is included) to June 1st at large $v_{\rm min}$.~For these interactions $A_1$ changes sign at small $v_{\rm min}$.~For the remaining interactions, the curve $t_{\rm max}(v_{\rm min})$ is flat and $A_1$ has the same sign for all $v_{\rm min}$ values.~Fig.~\ref{fig:tmax}, left panel, illustrates our results on $t_{\rm max}(v_{\rm min})$  focusing on the $\mathcal{O}_1$ and $\mathcal{O}_5$ interactions.
\begin{figure}[t]
\begin{center}
\includegraphics[width=\textwidth]{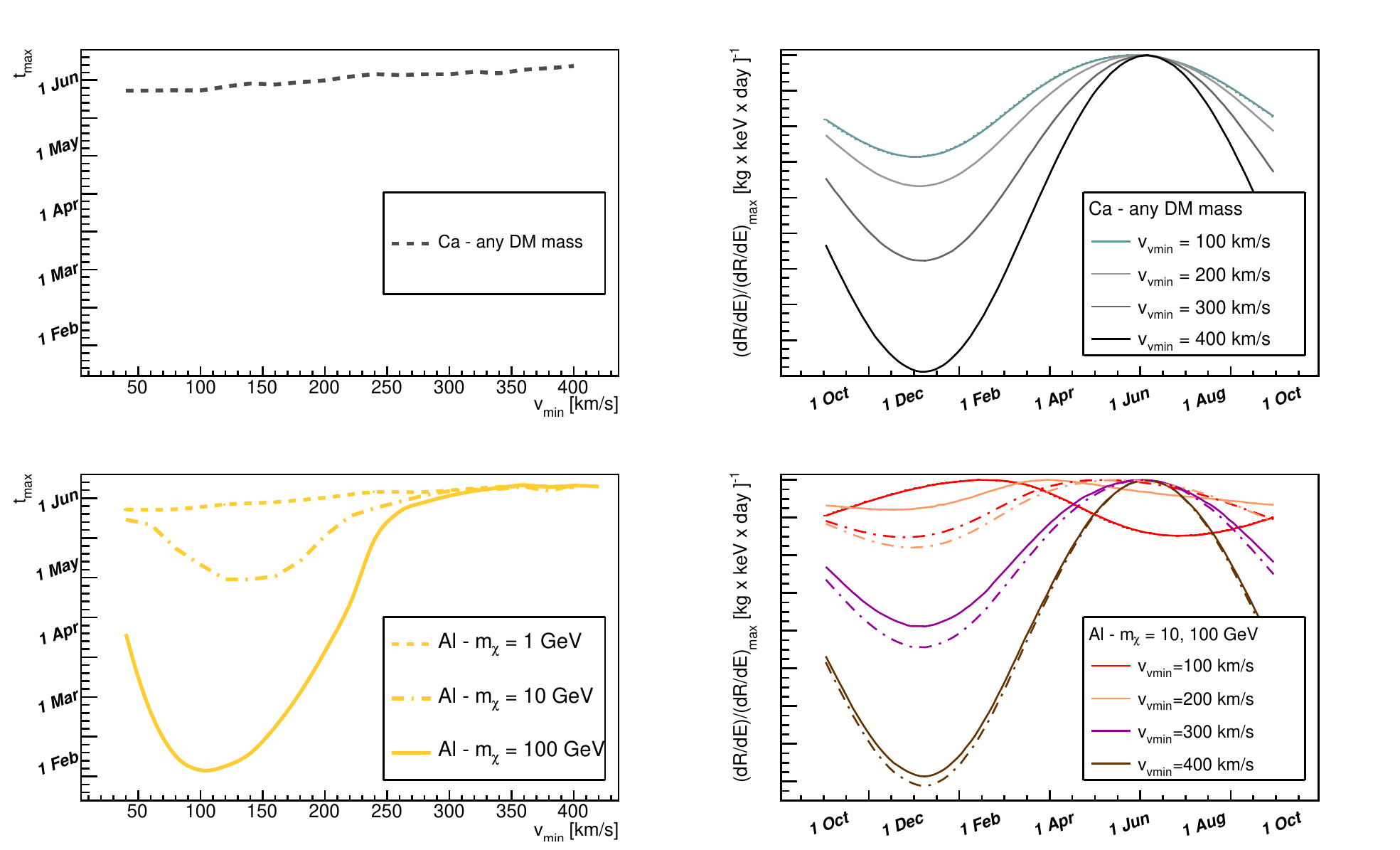}
\end{center}
\caption{{\it Left panels}.~Time of maximum differential rate of nuclear recoil events, $t_{\rm max}$, as a function of $v_{\rm min}$ for the DM magnetic dipole interaction model.~The top panel assumes calcium as a target material, whereas the bottom panel refers to aluminium.~{\it Right panels}.~Differential rate at different $v_{\rm min}$ values for calcium (top panel) and aluminium (bottom panel).~We set $m_\chi$ to the values in the legends.}  
\label{fig:target}
\end{figure}

Let us now focus on the DM magnetic dipole interaction model.~Here, we extend the results of~\cite{DelNobile:2015tza} to aluminium, calcium and oxygen, as these targets are of interest for CRESST detectors.~Fig.~\ref{fig:target} shows the $t_{\rm max}(v_{\rm min})$ curve for calcium (top panels) and aluminium (bottom panels).~The corresponding curve for oxygen is identical to the one associated with calcium.~The right panels in Fig.~\ref{fig:target} report differential nuclear recoil rates as a function of time for different values of $v_{\rm min}$ and, in the case of aluminium, for different DM particle masses.~In the case of calcium, the curve $t_{\rm max}(v_{\rm min})$ is approximately flat, as the term proportional to the $\tilde{\eta}$ function in Eq.~(\ref{eq:rate2}) dominates over the term proportional to $\eta$.~This is due to the fact that $^{16}$O and $^{40}$Ca have spin zero, which implies $W_{\Sigma'}=0$.~As a result, the surviving term in Eq.~(\ref{eq:mddm}) is proportional to $W^{pp}_M$ for both the $\eta$ and the $\tilde{\eta}$ contributions and the coefficient of $\tilde{\eta}$ is proportional to $1/v_{min}^{2}$, which explains why the $\tilde{\eta}$ contribution dominates over the $\eta$ one at small $v_{min}$..~In contrast, the curve $t_{\rm max}(v_{\rm min})$ has a minimum above 100~km~s$^{-1}$ in the case of aluminium.~This arises from the interplay of the $\eta$ and $\tilde{\eta}$ terms in Eq.~(\ref{eq:rate2}).~However, this minimum is pronounced only for $m_\chi>10$~GeV.~Therefore, for magnetic dipole DM-nucleus interactions, aluminium, calcium and  oxygen are characterised by the same, approximately flat $t_{\rm max}(v_{\rm min})$ curve for small $m_{\chi}$ values.~On the other hand, for the familiar spin-independent interaction, the $t_{\rm max}(v_{\rm min})$ curve span a period of about five months.~Consequently, it is reasonable to expect that timing information on the observed nuclear recoils can be used to discriminate spin-independent from magnetic dipole DM-nucleus interactions.~Since the phenomenology of the two models significantly differs in the small $v_{\rm min}$ limit, low-threshold experiments should then be optimal to discriminate between them.~While we here focused on the DM magnetic dipole, we verified that the same conclusions apply to the anapole DM model, which predicts a similar ${\rm d}\sigma_T/{\rm d} E_R$.


\section{Statistical framework}
\label{sec:stat}
In this section, we introduce the statistical framework used to assess the prospects for DM signal discovery and model selection via timing (and energy) information in a low-threshold experiment.~As anticipated, our goal is to determine whether a time dependent analysis can improve the sensitivity of CRESST in the presence of an unidentified background resembling the low-energy events recently reported by the CRESST collaboration.~Here, we assume that $n$ nuclear recoil events have been observed in a low-threshold detector.~The data sample is distributed over a signal region, $\Delta S$, that we divide into $N$ bins.~$\Delta S$ and the $N$ bins are one-dimensional energy intervals, when we only analyse the energy spectrum of the nuclear recoils (1D analysis).~They are two-dimensional, energy$\times$time intervals, when we assume that timing information on the nuclear recoils is also accessible (2D analysis).

\subsection{Signal discovery}
\label{sec:statsd}
Focusing on DM signal discovery, we compare the background-only hypothesis with an alternative, signal-plus-background hypothesis.~In the  background-only hypothesis the nuclear recoil events observed in the detector can be explained in terms of experimental backgrounds alone.~In the signal-plus-background hypothesis, the observed events include a DM signal contribution.~We model the energy spectrum of the background events by assuming that they are distributed in energy as the energy spectrum observed in CRESST-III, which presents an excess below 200~eV, as reported in~\cite{Abdelhameed:2019hmk} and described in the introduction.~We also assume that the background events are homogeneously distributed in time.~In our calculations, the number of background events, $\lambda_b$, in a pre-defined signal region $\Delta S$ is an input.~We set $\lambda_b$ to the desired value by varying the overall background normalisation, $\mathscr{B}$.~This model for the experimental background is only taken as a reference for the order of magnitude of the expected background level, as the energy spectrum reported in~\cite{Abdelhameed:2019hmk} was observed using CaWO$_4$, whereas here we consider Al$_2$O$_3$ as target material.~Specifically, for the experimental background we assume the energy spectrum~\cite{schmiedmayer2019calculation},
\begin{align}
\frac{{\rm d}N_b}{{\rm d}E_R} = \mathscr{B} \left( p_0 +p_1 E_R + p_2 e^{-E_R/p_3} \right)\,,
\label{eq:back}
\end{align}
where we set $p_0$, $p_1$, $p_2$ and $p_3$ to the best fit values we found by fitting Eq.~(\ref{eq:back}) with $\mathscr{B}=1$ to the data in~\cite{Abdelhameed:2019mac} (file `C3P1\_DetA\_full.dat'), namely:~$p_0=28$~keV$^{-1}$, $p_1=-0.8$~keV$^{-2}$, $p_2=19776$~keV$^{-1}$ and $p_3=0.0423$~keV.~In the signal-plus-background hypothesis, we model the DM signal contribution in terms of spin-independent interactions.~Specifically, we set the DM-nucleon scattering cross section, the proton to neutron coupling ratio and the DM particle mass to two benchmark set of values:~1) the best fit values found in a fit of the latest DAMA results~\cite{Bernabei:2018jrt}, namely~$\sigma_{\rm SI}=2.67\times10^{-38}$~cm$^2$, $r=c_p/c_n=-0.76$ and $m_\chi=11.17$~GeV, respectively~\cite{Kang:2018qvz}; and~2)~$\sigma_{\rm SI}=4\times10^{-42}$~cm$^2$, $r=c_p/c_n=1$ and $m_\chi=3$~GeV, respectively.~The first benchmark allows us to test the relative impact of energy and timing information on the discovery of a signal resembling the one reported by DAMA.~The second one corresponds to the 90\%~C.L. upper bound on the DM-nucleon scattering cross section reported by the DarkSide-50 collaboration for $m_\chi=3$~GeV~\cite{Agnes:2018ves}.~Notice that $m_\chi=3$~GeV is the smallest DM mass that can induce values of $v_{\rm min}$ below $200$~km~s$^{-1}$ for recoil energies below $\simeq 200$~eV, and that below $200$~km~s$^{-1}$ the modulation amplitude changes sign in the case of spin-independent interactions.

Let us now denote by $s_i$ the expected number of DM signal events in the $i$-th bin when the model parameters are set at one of the above benchmark points.~The expected number of signal events in the signal region $\Delta S$ is hence $\lambda_s=\sum_i s_i$.~Here, we adopt the notation of~\cite{Cowan:2010js} and introduce the signal strength parameter, $\mu$.~For $\mu=0$ there is no DM signal contribution to the observed nuclear recoil events, while for $\mu=1$ the DM contribution to the $i$-th bin is $s_i$.~Furthermore, we denote by $b_i$ the expected number of background events in the $i$-th bin and introduce the parameter $\theta$ (see Eq.~(\ref{eq:likelihood})) to model the uncertainties in the background amplitude.~Both in the 1D analysis and in the 2D analysis, we test the background-only hypothesis against the signal-plus-background hypothesis by means of the test statistic~\cite{Cowan:2010js},
\begin{align}
q_0 = \left \{ 
\begin{array}{ll} - 2 \ln \lambda  &\quad \hat{\mu}\ge0 \\
0  & \quad \hat{\mu}<0
\end{array}
\right.
\label{eq:lr}
\end{align}
where 
\begin{align}
\lambda=\frac{\mathscr{L}(0,\hat{\hat{\theta}})}{\mathscr{L}(\hat{\mu},\hat{\theta})}\,,
\end{align}
and $\mathscr{L}(\mu,\theta)$ is the likelihood function, 
\begin{align}
\mathscr{L}(\mu,\theta) = \prod_{i=1}^N \frac{(\mu s_i + \theta b_i)^{n_i}}{n_i!} e^{-(\mu s_i + \theta b_i)} \,. 
\label{eq:likelihood}
\end{align}
Here, $\hat{\hat{\theta}}$ is the value of $\theta$ that maximises $\mathscr{L}$ when $\mu=0$, while $(\hat{\mu},\hat{\theta})$ is the maximum likelihood estimator.~We sample the total number of events (background plus signal) in the $i$-th bin, $n_i$, from a Poisson distribution of mean $\bar{\mu}s_i+\bar{\theta}b_i$, where $\bar{\mu}$ and $\bar{\theta}$ are the hypothesised values for $\mu$ and $\theta$, respectively.~For example, when sampling $n_i$ under the background-only hypothesis, $\bar{\mu}=0$ and $\bar{\theta}=1$.~When sampling $n_i$ under the signal-plus-background hypothesis, $\bar{\mu}=1$ and  $\bar{\theta}=1$.~By repeatedly sampling $n_i$ under the background-only hypothesis, we obtain the probability density function (pdf) $f(q_0|0)$.~Sampling $n_i$ under the signal-plus-background hypothesis we obtain the pdf $f(q_0|1)$.~From $f(q_0|0)$ and $f(q_0|1)$, we obtain the $p$-value for signal discovery,
\begin{align}
p = \int_{q_{0}^{\rm med}}^{\infty} {\rm d}q_0\, f(q_0|0) \,,
\label{eq:p0}
\end{align}
where $q_{0}^{\rm med}$ is the median of $f(q_0|1)$.~In order to test the importance of the nuisance parameter $\theta$, we also evaluate the $p$-value in Eq.~(\ref{eq:p0}) for $\theta$ set to the constant value $\theta=\bar{\theta}=1$, so that $\hat{\hat{\theta}}=\hat{\theta}=1$.

\subsection{Model selection}
Focusing on DM model selection, we compare a ``null-hypothesis'' where DM couples to nuclei through the familiar spin-independent interaction, with an alternative hypothesis where DM couples to nuclei via a magnetic dipole moment.~Within the null-hypothesis, we model the DM contribution to the observed nuclear recoils by assuming $m_\chi=3$~GeV, $r=c_p/c_n=1$ and $\sigma_{\rm SI}=4\times 10^{-42}$~cm$^2$, as in one of the realisations of the signal-plus-background hypothesis considered in Sec.~\ref{sec:statsd}.~This set of parameters predicts 2.2 counts in an experimental setup consisting of an Al$_2$O$_3$ detector with an exposure of $23$~g$\times$year, when considering the interval $[0.012 - 5.4]$~keV as energy window.~Within the alternative hypothesis, we assume $m_\chi=3$~GeV and couplings set to $\sigma_{\rm MD}\equiv4 \alpha^2\lambda_\chi^2 =4.72\times 10^{-41}$~cm$^2$.~This reference cross section value predicts 2.2 counts for the above experimental setup.~Within both hypotheses, we model the experimental background as in Sec.~\ref{sec:statsd}.~Finally, we compare the two hypotheses by using the log-likelihood ratio test statistic,
\begin{align}
q =  - 2 \ln \widetilde{\lambda} \end{align}
where
\begin{align}
\widetilde{\lambda}=\frac{\mathscr{L}^{(0)}(\hat{\mu},\hat{\theta})}{\mathscr{L}^{(a)}(\hat{\mu},\hat{\theta})}
\end{align}
and
\begin{align}
\mathscr{L}^{(0)}(\mu,\theta) &= \prod_{i=1}^N \frac{(\mu s^0_i + \theta b_i)^{n_i}}{n_i!} e^{-(\mu s^0_i + \theta b_i)}\,, \nonumber\\
\mathscr{L}^{(a)}(\mu,\theta) &= \prod_{i=1}^N \frac{(\mu s^a_i + \theta b_i)^{n_i}}{n_i!} e^{-(\mu s^a_i + \theta b_i)}\,.
\label{eq:like0a}
\end{align}
Here, $s^0_i$ ($ s^a_i$) is the number of expected signal events in the $i$-th bin under the null-hypothesis (alternative hypothesis).~Importantly, $n_i$ denotes the same dataset in both lines of Eq.~(\ref{eq:like0a}). Analogously to the case of DM signal discovery, we sample $n_i$ from a Poisson distribution of mean $\bar{\mu}\,\bar{s}_i+\bar{\theta}b_i$.~Sampling $n_i$ under the null hypothesis, i.e.~$\bar{\mu}=1$, $\bar{\theta}=1$ and $\bar{s}_i=s_i^0$, we obtain the pdf $f_0(q|1)$.~Similarly, we obtain the pdf $f_a(q|1)$ when sampling $n_i$ under the alternative hypothesis, i.e.~$\bar{\mu}=1$, $\bar{\theta}=1$ and $\bar{s}_i=s_i^a$.~Here, we quantify the ability of a low-threshold experiment to reject the null in favour of the alternative hypothesis in terms of the $p$-value,
\begin{align}
p = \int_{q^{\rm med}}^{\infty} {\rm d}q\, f_0(q|1)\,,
\label{eq:p}
\end{align}
where $q^{\rm med}$ is the median of the pdf $f_a(q|1)$.

\begin{figure}[t]
\begin{center}
\includegraphics[width=\textwidth]{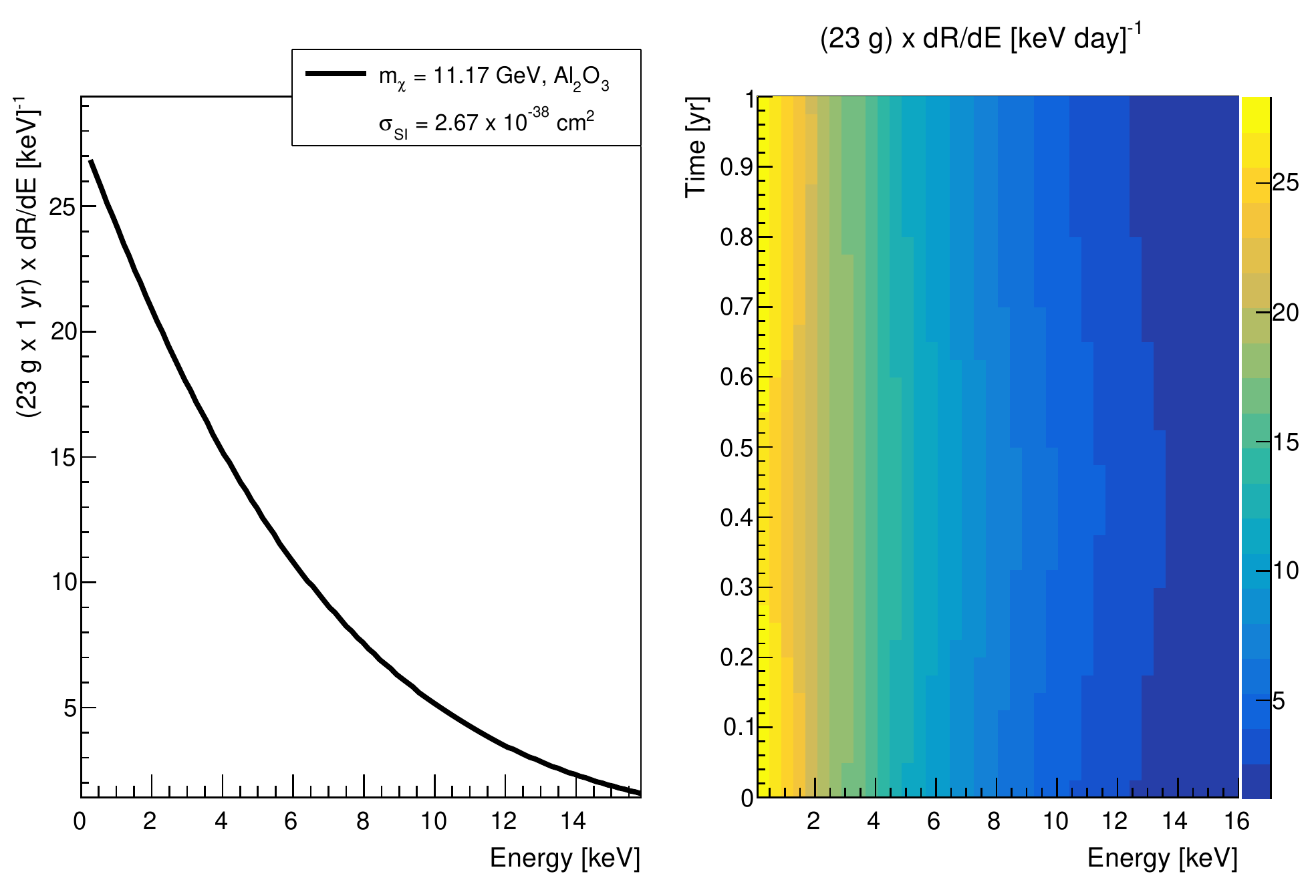}
\end{center}
\caption{Predicted energy (left panel) and energy$\times$time (right panel) distributions of DM-induced nuclear recoil events for our benchmark 1 scenario:~$\sigma_{\rm SI}=2.67\times10^{-38}$~cm$^2$, $r=c_p/c_n=-0.76$ and $m_\chi=11.17$~GeV.}  
\label{fig:signal_b1}
\end{figure}

\begin{figure}[t]
\begin{center}
\includegraphics[width=\textwidth]{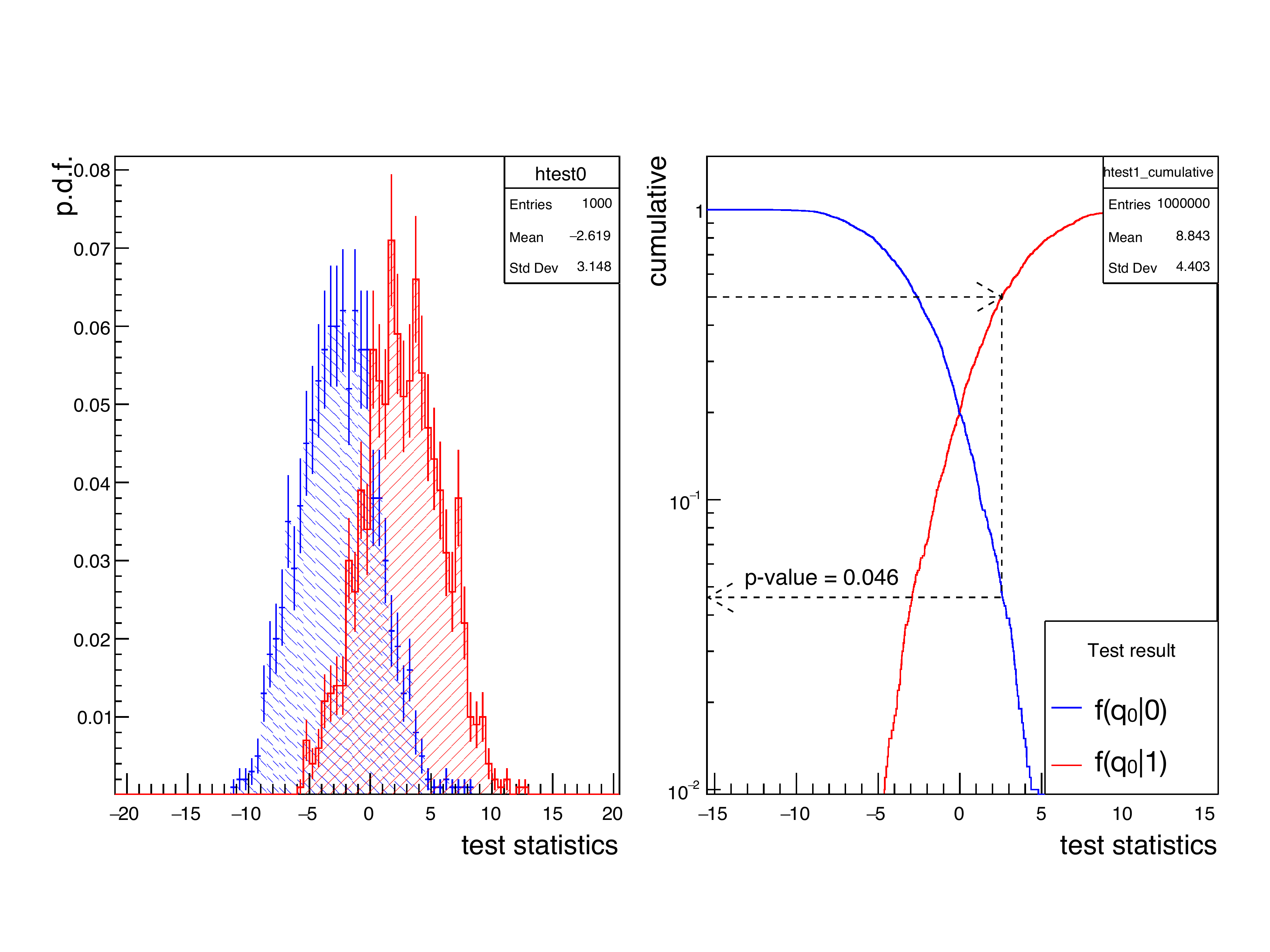}
\end{center}
\caption{Probability density functions $f(q_0|0)$ and $f(q_0|1)$ (left panel) and associated cumulative probability density functions (right panel) for a simulation assuming no timing information, $\lambda_b=10^3$ and our benchmark scenario 1.~The $p$-value for DM discovery associated with this specific example is $p=0.046$.}  
\label{fig:pdf_b1}
\end{figure}

\section{Results}
By applying the statistical methods of Sec.~\ref{sec:stat} to the theoretical framework of Sec.~\ref{sec:theory}, we now report the result of our $p$-value calculations for DM signal discovery, Sec.~\ref{sec:signal}, and model selection, Sec.~\ref{sec:model}.~Here, we assume that the number of successes in observing a value of $q$ ($q_0$)  above $q^{\rm med}$ ($q^{\rm med}_0$) obeys a binomial distribution of variance $\sigma^2=p \mathscr{N}(1-p)$, where $\mathscr{N}$ is the number of times we Monte Carlo generate the sample $n_i$.~Consequently, the upper bound on the population error of our $p$-value estimates is given by $1/\sqrt{4\mathscr{N}}$.~Since we choose $p=0.1$ (i.e.~90\% C.L.) as a reference $p$-value, we set $\mathscr{N}$ to $10^3$.~In the tables presented below, $p\simeq 0$ indicates that the result of our calculations is comparable or smaller than the associated error, $1/\sqrt{4\mathscr{N}}\simeq 0.016$.

\subsection{Signal discovery}
\label{sec:signal}
We start by testing the background-only hypothesis against the signal-plus-background hypothesis. We model the experimental background via Eq.~(\ref{eq:back}) and the DM contribution to the observed nuclear recoils in terms of spin-independent interactions.~Here, we set the DM-nucleon scattering cross section, the proton to neutron coupling ratio and the DM particle mass to:
\begin{enumerate}
\item $\sigma_{\rm SI}=2.67\times10^{-38}$~cm$^2$, $r=c_p/c_n=-0.76$ and $m_\chi=11.17$~GeV, respectively~\cite{Kang:2018qvz}.~Furthermore, $\Delta S =$ [105~eV, 16~keV].
\item $\sigma_{\rm SI}=4\times10^{-42}$~cm$^2$, $r_p/r_n=1$ and $m_\chi=3$~GeV, respectively~\cite{Agnes:2018ves}.~In addition, $\Delta S =$ [12~eV, 5.4~keV].
\end{enumerate}
Our motivations for focusing on these two benchmarks have been discussed in Sec.~\ref{sec:timing}.~The first one corresponds to the best fit to the latest DAMA results, the second one to the 90\% C.L. upper limit on the DM-nucleon scattering cross section reported by the DarkSide-50 experiment.~In both cases, we assume Al$_{2}$O$_3$ as target material and a minimum  exposure of 23~g$\times$158~days.~In the 1D analyses nuclear recoils are distributed in 10 energy bins, while in the 2D analyses they are distributed in 10$^2$ energy$\times$time bins.~We test the results using both linear and logarithmic binning size.~We use the test statistic in Eq.~(\ref{eq:lr}).

\subsubsection{Benchmark 1}
Focusing on the first benchmark scenario, we now present our results on the ability of a low-threshold experiment to reject the background-only hypothesis.~For this scenario, Fig.~\ref{fig:signal_b1} shows the predicted energy (left panel) and energy$\times$time (right panel) distributions of DM-induced nuclear recoil events.~As expected from our analysis of $A_1(v_{\rm min})$ and $t_{\rm max}(v_{\rm min})$ in Sec.~\ref{sec:timing}, the 2D distribution peaks around December/January at small energies and in June at high energies.

Using the information contained in Fig.~\ref{fig:signal_b1}, we first test the ability of a low-threshold experiment to reject the background-only hypothesis when the background parameter $\theta$ is set to the constant value $\theta=\bar{\theta}=1$.~We will treat $\theta$ as a nuisance parameter later on.~As an illustrative example of the geometric interpretation of the $p$-value for signal discovery in Eq.~(\ref{eq:p}), Fig.~\ref{fig:pdf_b1} shows the pdfs $f(q_0|0)$ and $f(q_0|1)$ and the associated cumulative pdfs for a simulation assuming no timing information, $\lambda_b=10^3$, $\lambda_s\simeq141$ and $\Delta S =$ [105~eV, 16~keV] as signal region.~As reported in the legend, the $p$-value associated with this specific example is $p=0.046$.

\begin{table}[t]
   \centering
   \begin{tabular}{ccccccc} 
          \hline
mass & time  & $\lambda_b$ & $\lambda_{s}$ &N inter& p-value 1D   & p-value 2D  \\ \hline
23 g  & 1 yr     & $10^3$     &    $\simeq141$  & $10^3$    & $\simeq 0$ &   $\simeq 0$  \\
 $''$   & $''$     & $10^4$     &     $''$                 & $''$          & 0.046          & 0.045   \\
 $''$   & $''$     & $10^5$     &     $''$                 & $''$          &  0.284     &  0.247 \\
 2\;x\;23g   & $''$     &  $''$      &     $283$                 & $''$          &  0.161     &  0.163 \\
 3\;x\;23g   & $''$     &  $''$      &     $424$                 & $''$          &  0.055     &  0.077 \\
 \hline
 \end{tabular}
   \caption{$p$-values for DM signal discovery in the case of a DAMA-like signal (benchmark scenario 1), known background and no nuisance parameters.~The number of bins is set to 10 both for energy and time.~The first and second column correspond to detector mass and time of data taking, respectively.~The fifth column gives the number of independent test statistic evaluations in our Monte Carlo simulations.~The $p$-values that we obtained in 1D (2D) analyses are reported in the sixth (seventh) column.}
   \label{tab:b1}
\end{table}

\begin{table}[t]
   \centering
   \begin{tabular}{ccccccc} 
          \hline 
 mass & time                & $\lambda_b$        & $\lambda_{s}$  &N inter     & p-value 1D                         & p-value 2D  \\ \hline
23 g          & 158 days        & $500$                   & $\simeq 69$      & $10^3$    & 0.086     & 0.083  \\
$''$            & 1 yr                 & $10^3$                 & $\simeq159$     &  $''$   & $\simeq 0$      & $\simeq0$  \\
 $''$           &  $''$                    & $10^4$                 &     $''$                    & $''$      &  0.21          & 0.212\\
 $''$           &   $''$                    & $10^5$                 &   $''$                      &     $''$                   &  0.398           &  0.415   \\
23 g      & 2 yr                  & $2\cdot10^3$       & $\simeq319$     &  $''$  & $ \simeq 0$   & $\simeq 0$  \\
 $''$           &   $''$                    & $2\cdot10^4$       &  $''$                        & $    ''    $                 & 0.127                        & 0.14 \\
  $''$           &    $''$                   & $2\cdot10^5$       &  $''$                        & $    ''    $                  & 0.34                        & 0.362  \\
 \hline
 \end{tabular}
   \caption{Same as Tab.~\ref{tab:b1}, but now assuming an unknown background amplitude as a nuisance parameter.}
   \label{tab:b1_nuis}
\end{table}

Tab.~\ref{tab:b1} extends the results in Fig.~\ref{fig:pdf_b1} and shows the $p$-values we find when $\theta$ is set to the constant value $\theta=\bar{\theta}=1$ for different combinations of $\lambda_b$, the expected number of background events, and experimental exposures.~More specifically, we consider values of $\lambda_b$ ranging from $10^3$ to $10^5$ and experimental exposures varying from 23~g$\times$year to $3\times$23~g$\times$year.~Importantly, in Tab.~\ref{tab:b1} we report $p$-values for signal discovery that we obtain when timing information is taken into account (2D analysis) and when timing information is not available (1D analysis).~As one can see from Tab.~\ref{tab:b1}, timing information has a marginal impact on the ability of a low-threshold experiment to reject the background-only hypothesis when the background events are modelled as in Eq.~(\ref{eq:back}) and the observed DM signal resembles the one reported by DAMA.

We now present our results on the ability of a low-threshold experiment to reject the background-only hypothesis when the background parameter $\theta$ is considered as a nuisance parameter.~Tab.~\ref{tab:b1_nuis} shows the $p$-values for signal discovery that we find in simulations where $\lambda_b$ varies from $500$ to $2\times10^5$, the experimental exposure varies from 23~g$\times$158~days (corresponding to the exposure reached in the CRESST-III analysis~\cite{Abdelhameed:2019hmk}) to 23~g$\times$2~years and $\theta$ is treated as a nuisance parameter.~As in the case of Tab.~$\ref{tab:b1}$, Tab.~\ref{tab:b1_nuis} shows the results we find when timing information on the observed nuclear recoils is taken into account, and when the latter is not available.~Comparing Tab.~\ref{tab:b1} with~\ref{tab:b1_nuis}, one can see that in all cases the $p$-values we find when $\theta$ is a nuisance parameters are larger than when the experimental background is assumed to be known.~This result is expected, as the Neyman-Pearson lemma (see~\cite{Cowan:2010js} and references therein) states that the likelihood ratio in Eq.~(\ref{eq:lr}) for $\theta=\bar{\theta}=1$ is the test statistic with the highest significance power, i.e.~the highest probability of rejecting the background-only hypothesis if the background-plus-signal hypothesis is true.~Tab.~\ref{tab:b1_nuis} also shows that timing information has a marginal impact on the ability of a low-threshold experiment to reject the background-only hypothesis in favour of benchmark 1 in the case of an unknown background amplitude.

\begin{table}[t]
   \centering
   \begin{tabular}{ccccccc} 
          \hline
  mass & time           & $\lambda_{s}$        & $\lambda_{b}$   & N iter     & p-value 1D & p-value 2d  \\ \hline
23 g               &158 days    & $\simeq 1$         & 500                 &    $10^3$                  & 0.436   &  0.482\\
 230 g           &$''$              &$\simeq 22$              & $10^4$                & $''$     & 0.323     &  0.346 \\
$''$                &2 yr             &$\simeq 45$                   & $2\cdot 10^4$     &  $''$     & 0.223    &  0.263   \\
$''$                &5 yr              & $\simeq 112$   & $5\cdot 10^4$      &   $ ''$           & 0.151     &   0.151  \\ 
 1 kg            &$''$               & $\simeq 491$   &  $''$                    &   $''$                  & $\simeq 0$   &    $\simeq 0$\\
$''$               &$''$               & $''$ &$5\cdot 10^5$    &$ ''$          & $0.065$      &   $0.094$\\
 \hline
 \end{tabular}
   \caption{\small{Same as Tab.~\ref{tab:b1_nuis}, but now modelling the DM-nucleon interaction as in our benchmark 2 scenario.}}
   \label{tab:b2_nuis}
\end{table}

\subsubsection{Benchmark 2}
Now we turn our attention to the second benchmark scenario considered in this investigation, where  $\sigma_{\rm SI}=4\times10^{-42}$~cm$^2$, $r=c_p/c_n=1$ and $m_\chi=3$~GeV.~Focusing on this second scenario, below we present our results on the ability of a low-threshold experiment to reject the background-only hypothesis.~We only consider the case of a background of unknown amplitude and treat $\theta$ as a nuisance parameter.~Tab.~\ref{tab:b2_nuis} shows the $p$-values for signal discovery that we find comparing the background-only hypothesis with a DM signal model corresponding to our benchmark 2.~In our analysis of benchmark 2, we let the number of background events, $\lambda_b$, vary from $500$ to $5\times10^5$ and consider experimental exposures varying from 23~g$\times$158~days to 1~kg$\times$5~years.

Our results for benchmark 2 are summarised in Tab.~\ref{tab:b2_nuis}.~In the 1D analysis, the $p$-values we obtain range from about 0.44, for $\lambda_b=500$ and an exposure of 23~g$\times$158~days, to about 0.06, for $\lambda_b=5\times10^5$ and an exposure of 1~kg$\times$5~years.~Consistently with benchmark 2 being about two orders of magnitude below the current CRESST-III exclusion limit on  $\sigma_{\rm SI}$ for $m_\chi=3$~GeV, we find that the CRESST experiment must operate with an exposure about 100 larger than the the current one, i.e.~23~g$\times$158~days, in order to achieve a 90\%~C.L. discovery of a DM model like benchmark 2.~Remarkably, in all cases investigated here we find comparable $p$-values in the 1D  analysis and in the  2D analysis, where timing information is taken into account.~We therefore conclude that timing information has a marginal impact on the ability of a low-threshold experiment to reject the background-only hypothesis in favour of benchmark 2.~These results have been verified to be stable both using the linear and the logarithmic distributed binning size.

\subsection{Model selection}
\label{sec:model}
We conclude this section by presenting our results on the ability of a low-threshold experiment to reject spin-independent in favour of magnetic dipole DM-nucleon interactions.~We refer to the former case as ``null-hypothesis'' and to the latter as ``alternative hypothesis''.~As anticipated, we model the null-hypothesis by assuming $m_\chi=3$~GeV, $r_p/r_n=1$ and $\sigma_{\rm SI}=4\times 10^{-42}$~cm$^2$.~This set of parameters predicts 2.2 counts in the $[0.012 - 5.4]$~keV energy window for an Al$_2$O$_3$ detector with an exposure of $23$~g$\times$year.~We model the alternative hypothesis by assuming $m_\chi=3$~GeV and a cross section $\sigma_{\rm MD}\equiv4 \alpha \lambda_\chi^2 =4.72\times 10^{-41}$~cm$^2$.~Consistently, this reference cross section value produces 2.2 counts in the same Al$_2$O$_3$ detector and energy window.

Tab.~\ref{tab:ms} shows the $p$-values we find when comparing null and alternative hypotheses.~In this investigation, we consider experimental exposures ranging from 23~g$\times$158~days to 1~kg$\times$5~years, and $\lambda_b$ varying from $10^3$ to $2\times10^4$.~We report both $p$-values for the 1D analysis and $p$-values for the 2D analysis, where timing information is taken into account.~In all cases, we treat the background parameter $\theta$ as a nuisance parameter.~The information in Tab.~\ref{tab:ms} is two-fold.~On the one hand we find that a 2D-analysis based on the energy and time distribution of the event rate produces more significant results with respect to the 1D analysis, at equal exposure. While previous results do not depend on the linear or logarithmic binning method, the different significance of the 2D and 1D analyses in this case strictly depends on the better accuracy of the logarithmic method. ~On the other hand, our results show that a 90\% C.L. rejection of spin-independent interactions in favour of a DM magnetic dipole coupling is feasible at the CRESST experiment with an exposure of 460~g$\times$year and $2\times10^3$ background events.



\begin{table}[h!]
   \centering
   \begin{tabular}{ccccccc} 
          \hline
  mass & time  & $\lambda_b$ & $\lambda_{s}$ &N inter& p-value 1D & p-value  2D  \\ \hline
 23~g  &  1 yr     & $10^3$             &     $\simeq  2.2 $     & $10^3$    & 0.486 &  0.442  \\
  230~g &  2 yr     & $2\cdot 10^3$      &     $\simeq  44 $    & $''$           & 0.107 &  0.065  \\
   $''$  &  5 yr      & $5 \cdot 10^3$           &     $\simeq 110 $      & $''$           & 0.024 &    0.005 ($\simeq 0$)\\
 1 kg        &  $''$     & $2 \cdot 10^4$           &     $\simeq 478 $      & $''$           & $\simeq 0$ &   $ \simeq 0$ \\
 \hline
 \end{tabular}
   \caption{$p$-values for rejecting spin-independent interactions with $m_\chi=3$~GeV and $\sigma_{\rm SI}=4\times 10^{-42}$~cm$^2$ in favour of a magnetic dipole coupling producing the same number of events in $\Delta S =$ [12~eV, 5.4~keV].~We report results with and without timing information, obtained with logarithmic binning.~The notation is the same as in previous tables.}
   \label{tab:ms}
\end{table}

\section{Summary and conclusions}
\label{sec:conclusion}
We assessed the prospects for DM signal discovery and model selection via timing information in a low-threshold experiment.~In all calculations, we assumed a time-independent experimental background with an energy spectrum resembling the one of the low-energy events observed in~\cite{Abdelhameed:2019hmk}.~We focused on Al$_2$O$_3$ as a detector material, as it is composed of two relatively light elements (oxygen and aluminium) and showed very good performances, with energy thresholds as low as 19 eV~\cite{Angloher:2017sxg}.

As a preparatory step, we critically reviewed the timing information that is available to low-threshold multi-target experiments, focusing on interaction-dependent features, such as the shape of the $t_{\rm max}(v_{\rm min})$ curve, which can depend on the target material.~We also explored the significance of higher-order harmonics in a Fourier series expansion of the predicted nuclear recoil rate.~We discussed these aspects within the general non-relativistic effective theory of DM-nucleon interactions, and for specific interaction models, such as the DM magnetic dipole coupling.

Focusing on DM signal discovery, we compared a background-only hypothesis with an alternative, background-plus-signal hypothesis.~Here, we modelled the interaction between DM and nuclei in terms of spin-independent interactions and considered two benchmark scenarios separately.~The first scenario corresponds to the best fit mass and couplings to the latest DAMA/LIBRA results.~The second one corresponds to the 90\% C.L upper limit on the spin-independent DM-nucleon scattering cross section reported by DarkSide-50 for a DM particle mass of 3 GeV.~Focusing on DM model selection, we compared a ``null-hypothesis'' where DM couples to nuclei through the familiar spin-independent interaction (and parameters as in the second scenario described above) with an alternative hypothesis where DM couples to nuclei via a magnetic dipole moment.~We presented our results on the ability of a low-threshold experiment to discover a signal above a background resembling the excess found in the CRESST-III energy spectrum, or reject the spin-independent interaction in favour of a magnetic dipole coupling in terms of $p$-values, that we list in Tabs.~\ref{tab:b1}, \ref{tab:b1_nuis}, \ref{tab:b2_nuis} and \ref{tab:ms}.~Importantly, we performed our $p$-value calculations under two different assumptions:~1) Taking timing information into account.~2) Assuming that the latter is not available.

Remarkably, we found that while timing information has a marginal impact on the potential for DM signal discover, it can be beneficial for model selection in a low-threshold experiment.~This applies to all cases studied here and reported in Tabs.~\ref{tab:b1}, \ref{tab:b1_nuis}, \ref{tab:b2_nuis} and \ref{tab:ms}.~On the one hand, our results indicate that for this class of experiments it might not be worth tackling the challenge of operating cryogenic detectors under stable conditions for long periods, until the goal remains observing the first dark matter signal. On the other hand, if cryogenic experiments undertake such an effort, in case of detection of a positive dark matter signal, the timing information could provide more reliable results on the dark matter nature.~For example, we found that the $p$-value for rejecting spin-independent interactions in favour of a magnetic dipole coupling is of about 0.1 when the exposure is 460 g$\times$year, while about 0.06 if timing information is taken into accout.~Our simulations are limited to at most 5 years of data taking, specific benchmarks and to the assumption of a background constant in time; it should be considered as a starting point towards further, refined analyses of the impact of timing information on the performance of future low-threshold experiments. 

In addition, our results also show that for the set of model parameters considered here a 90\% C.L. rejection of spin-independent interactions in favour of a magnetic dipole coupling is feasible with an upgrade of the CRESST experiment~\cite{CRESST:2015djg}, where timing information is not still available; see Tab.~\ref{tab:ms}, second row.~In order to consolidate our findings, we compared spin-independent and magnetic dipole interactions under the assumption of a detector efficiency below 100\%.~Specifically, in our likelihood analysis we multiplied the expected number of signal events in each bin, $s_i$, by a detector efficiency sampled from a uniform distribution in the 50\% - 70\% range.~For an exposure of 1~kg$\times$2~years and a background level of $5\times10^{3}$, we found a $p$-value for model selection of 0.072.~As expected, the experimental sensitivity diminishes when detector efficiency is taken into account.~At the same time, we found that a 90\% C.L. rejection of spin-independent interactions in favour of a magnetic dipole coupling is feasible at a CRESST-III experiment, even when the detector efficiency is below 100\%.

\acknowledgments 
This work has been performed within a double doctoral degree agreement between Chalmers University and Technology and the Gran Sasso Science Institute.~The authors thank, Federica Petricca, Paolo Gorla, Karoline Sch\"affner and Francesco Vissani for stimulating and supporting the implementation of the agreement between the two institutes, and Florian Reindl and the CRESST Collaboration for the many valuable discussions and suggestions which inspired the research contained in this work.~RC acknowledges support from an individual research grant from the Swedish Research Council, dnr.~2018-05029.~The results presented here made use of the computer programmes \texttt{Wolfram Mathematica}~\cite{Mathematica} and \texttt{DMFormFactor}~\cite{Anand:2013yka}.

\appendix
\section{DM response functions}
\begin{eqnarray}
 R_{M}^{\tau \tau^\prime}\left(v_T^{\perp 2}, {q^2 \over m_N^2}\right) &=& c_1^\tau c_1^{\tau^\prime } + {J_\chi (J_\chi+1) \over 3} \left[ {q^2 \over m_N^2} v_T^{\perp 2} c_5^\tau c_5^{\tau^\prime } 
 + v_T^{\perp 2}c_8^\tau c_8^{\tau^\prime }
+ {q^2 \over m_N^2} c_{11}^\tau c_{11}^{\tau^\prime } \right] \nonumber \\
 R_{\Phi^{\prime \prime}}^{\tau \tau^\prime}\left(v_T^{\perp 2}, {q^2 \over m_N^2}\right) &=& {q^2 \over 4 m_N^2} c_3^\tau c_3^{\tau^\prime } + {J_\chi (J_\chi+1) \over 12} 
 \left( c_{12}^\tau-{q^2 \over m_N^2} c_{15}^\tau\right)  \left( c_{12}^{\tau^\prime }-{q^2 \over m_N^2}c_{15}^{\tau^\prime} \right)  \nonumber \\
 R_{\Phi^{\prime \prime} M}^{\tau \tau^\prime}\left(v_T^{\perp 2}, {q^2 \over m_N^2}\right) &=&  c_3^\tau c_1^{\tau^\prime } + {J_\chi (J_\chi+1) \over 3}  
 \left( c_{12}^\tau -{q^2 \over m_N^2} c_{15}^\tau \right) c_{11}^{\tau^\prime } \nonumber \\
  R_{\tilde{\Phi}^\prime}^{\tau \tau^\prime}\left(v_T^{\perp 2}, {q^2 \over m_N^2}\right) &=&{J_\chi (J_\chi+1) \over 12} \left[ c_{12}^\tau c_{12}^{\tau^\prime }+{q^2 \over m_N^2}  c_{13}^\tau c_{13}^{\tau^\prime}  \right] \nonumber \\
   R_{\Sigma^{\prime \prime}}^{\tau \tau^\prime}\left(v_T^{\perp 2}, {q^2 \over m_N^2}\right)  &=&{q^2 \over 4 m_N^2} c_{10}^\tau  c_{10}^{\tau^\prime } +
  {J_\chi (J_\chi+1) \over 12} \left[ c_4^\tau c_4^{\tau^\prime} 
+ {q^2 \over m_N^2} ( c_4^\tau c_6^{\tau^\prime }+c_6^\tau c_4^{\tau^\prime })+
 {q^4 \over m_N^4} c_{6}^\tau c_{6}^{\tau^\prime } \right. \nonumber \\
&+& \left. v_T^{\perp 2} c_{12}^\tau c_{12}^{\tau^\prime }+{q^2 \over m_N^2} v_T^{\perp 2} c_{13}^\tau c_{13}^{\tau^\prime } \right] \nonumber \\
    R_{\Sigma^\prime}^{\tau \tau^\prime}\left(v_T^{\perp 2}, {q^2 \over m_N^2}\right)  &=&{1 \over 8} \left[ {q^2 \over  m_N^2}  v_T^{\perp 2} c_{3}^\tau  c_{3}^{\tau^\prime } + v_T^{\perp 2}  c_{7}^\tau  c_{7}^{\tau^\prime }  \right]  
       {J_\chi (J_\chi+1) \over 12} \left[ c_4^\tau c_4^{\tau^\prime} + {q^2 \over m_N^2} c_9^\tau c_9^{\tau^\prime }  \right.\nonumber \\
       &+& \left. {v_T^{\perp 2} \over 2} \left(c_{12}^\tau-{q^2 \over m_N^2}c_{15}^\tau \right)  
       \left( c_{12}^{\tau^\prime }-{q^2 \over m_N^2}c_{15}^{\tau \prime} \right) 
       + {q^2 \over 2 m_N^2} v_T^{\perp 2}  c_{14}^\tau c_{14}^{\tau^\prime } \right] \nonumber \\
     R_{\Delta}^{\tau \tau^\prime}\left(v_T^{\perp 2}, {q^2 \over m_N^2}\right)&=&  {J_\chi (J_\chi+1) \over 3} \left[ {q^2 \over m_N^2} c_{5}^\tau c_{5}^{\tau^\prime }+ c_{8}^\tau c_{8}^{\tau^\prime } \right] \nonumber \\
 R_{\Delta \Sigma^\prime}^{\tau \tau^\prime}\left(v_T^{\perp 2}, {q^2 \over m_N^2}\right)&=& {J_\chi (J_\chi+1) \over 3} \left[c_{5}^\tau c_{4}^{\tau^\prime }-c_8^\tau c_9^{\tau^\prime} \right], \nonumber\\
 \label{eq:R}
\end{eqnarray}
where $J_\chi$ is the DM particle spin.~In all numerical application in the work, for DM we assume $J_\chi=1/2$.

\bibliographystyle{JHEP}
\bibliography{ref,ref2}

\end{document}